\documentclass[iop]{emulateapj}
\usepackage{ natbib}
\usepackage{xcolor}

\shorttitle{The Binary System AS\,386} \shortauthors{S.~Khokhlov, A.~Miroshnichenko, S.~Zharikov  et al.}

\begin{document}

\title{Toward Understanding the B[e] Phenomenon. VII.\\  AS\,386, a single-lined binary with a candidate black hole component.}

\author{S.~A.~Khokhlov$^{1,2,3}$}
\affil{$^1$NNLOT, Al-Farabi Kazakh National University, Al-Farabi Ave., 71, 050038, Almaty, Kazakhstan}
\affil{$^2$Fesenkov Astrophysical Institute, Observatory, 23, Almaty, 050020, Kazakhstan}
\affil{$^3$Al-Farabi Kazakh National University, Al-Farabi Ave., 71, 050040, Almaty, Kazakhstan}

\author{A.~S.~Miroshnichenko$^{4,1,2}$}
\affil{$^4$Department of Physics and Astronomy, University of North
Carolina at Greensboro, P.O. Box 26170, Greensboro, NC 27402--6170, USA}

\author{S.~V.~Zharikov$^5$}
\affil{$^5$Instituto de Astronomia, Universidad Nacional Aut\'onoma
de M\'exico, Apdo. Postal 877, Ensenada, 22800, Baja California, M\'exico}

\author{N.~Manset$^6$}
\affil{$^6$CFHT Corporation, Kamuela, HI 96743, USA}

\author{A.~A.~Arkharov$^7$, N.~Efimova$^7$, S.~Klimanov$^7$, V~M.~Larionov$^{7,8}$}
\affil{$^7$Pulkovo Astronomical Observatory of the Russian Academy of Sciences, St. Petersburg, 196140, Russia}
\affil{$^8$V.~V.~Sobolev Astronomical Institute of the St.-Petersburg State University, St.-Petersburg, 198504, Russia}

\author{A.~V.~Kusakin$^9$, R.~I.~Kokumbaeva$^9$, Ch.~T.~Omarov$^9$, K.~S.~Kuratov$^{1,3,9}$, A.~K.~Kuratova$^{1,3,9}$}
\affil{$^9$Fesenkov Astrophysical Institute, Observatory, 23, Almaty, 050020, Kazakhstan}

\author{R.~J.~Rudy$^{10}$, E.~A.~Laag$^{10}$, K.~B.~Crawford$^{10}$, T.~K.~Swift$^{10,11}$}
\affil{$^{10}$Aerospace Corporation, Los Angeles, CA 90009, USA}
\affil{$^{11}$Massachusetts Institute of Technology, 77 Massachusetts Ave, Cambridge, MA 02139, USA}

\author{R.~C.~Puetter$^{12}$}
\affil{$^{12}$University of California San Diego, 9500 Gilman Dr., La Jolla, CA 92093, USA}

\author{R.~B.~Perry$^{13}$}
\affil{$^{13}$ NASA Langley Research Center, Hampton, VA 23681, (Retired), USA}

\author{S. D. Chojnowski$^{14}$}
\affil{$^{14}$Apache Point Observatory and New Mexico State University, P.O. Box 59, Sunspot, NM, 88349--0059, USA}

\author{A.~Agishev$^3$}
\affil{$^3$Al-Farabi Kazakh National University, Al-Farabi Ave., 71, 050040, Almaty, Kazakhstan}

\author{D.~B.~Caton$^{15}$, R.~L.~Hawkins$^{15}$}
\affil{$^{15}$Dark Sky Observatory, Department of Physics and Astronomy, Appalachian State University, 525 Rivers St, Boone, NC 28608--2106, USA}

\author{A.~B.~Smith$^{16}$}
\affil{$^{16}$Gemini Observatory, Northern Operations Center, 670 N. A'ohoku Place, Hilo, Hawaii, 96720, USA}

\author{D.~E.~Reichart$^{17}$, V.~V.~Kouprianov$^{17,7}$, J.~B.~Haislip$^{17}$}
\affil{$^{17}$Department of Physics and Astronomy, University of North Carolina at Chapel Hill, Campus Box 3255, Chapel Hill, NC 27599, USA}

\begin{abstract}
We report the results of spectroscopic and photometric observations of the emission-line object AS\,386.
For the first time we found that it exhibits the B[e] phenomenon and fits the definition of an FS\,CMa type object.
The optical spectrum shows the presence of a B-type star with the following properties: T$_{\rm eff} = 11000\pm500$ K, $\log$ L/L$_{\odot} = 3.7\pm0.3$, a mass of $7\pm1$ M$_{\odot}$, and a distance $D = 2.4\pm0.3$ kpc from the Sun.
We detected regular radial velocity variations of both absorption and emission lines with the following orbital parameters: P$_{\rm orb} = 131.27\pm0.09$ days, semi-amplitude $K_{1} = 51.7\pm3.0$ km\,s$^{-1}$, systemic radial velocity $\gamma = -31.8\pm2.6$ km\,s$^{-1}$,  and a mass function of $f(m) = 1.9\pm0.3$ M$_{\odot}$.
AS\,386 exhibits irregular variations of the optical brightness  ($V=10.92\pm0.05$ mag), while the near-IR brightness varies up to $\sim 0.3$ mag following the spectroscopic period. We explain this behavior by a variable illumination of the dusty disk inner rim by the B-type component.
Doppler tomography based on the orbital variations of emission-line profiles shows that the material is distributed near the B-type component and in a circumbinary disk. We conclude that the system has undergone a strong mass transfer that created the circumstellar material and increased the B-type component mass. The absence of any traces of a secondary component, whose mass should be $\ge$ 7 M$_{\odot}$, suggests that it is most likely a black hole.
\end{abstract}


\section{Introduction} \label{intro}
A large number of B-type emission-line stars with infrared (IR) excess has been detected in various IR sky surveys by cross-identification with their optical positions. The excess is typically due to radiation of circumstellar gas (e.g., in Be stars and massive stars with strong stellar winds) and dust (e.g., objects with the B[e] phenomenon, which refers to the presence of emission lines, including forbidden, and IR excess due to dust radiation). Despite a strong progress in understanding of these complex objects in recent years, many of them are still poorly studied.

One group of objects with the B[e] phenomenon whose nature is not well understood is a recently defined group of FS\,CMa objects \citep{2007ApJ...667..497M}.
Most of its members show strong emission-line spectra, which cannot be explained by regular winds from single B-type stars.
A leading hypothesis about the nature of FS\,CMa objects is that they are binary systems after a mass transfer phase, which is responsible for most of the
circumstellar material including the dust. Over two dozens of $\sim$70 currently known group members and candidates have shown signatures of binarity
\citep[see][for a recent review]{2017ASPC..508..285M}. However, detecting secondary components FS\,CMa objects, where a B-star dominates radiation in the
optical region and circumstellar dust takes over in the IR, is not an easy task because of
several features, which include large brightness differences \citep[e.g., CI\,Cam,][]{2006ASPC..355..305B} and/or mass ratios \citep[e.g., MWC\,728,][]{2015ApJ...809..129M}.
Nevertheless, our ongoing program of spectroscopic and photometric monitoring of the group objects is being resulted in revealing new binaries among them thus giving
further support to the mentioned above hypothesis.

This paper is devoted to a study of AS\,386, a $V \sim$ 11.0 mag star located in Cygnus ($l = 75\fdg3, b = +2\fdg6$). The presence of the H$\alpha$ emission line in its spectrum was first reported by \citet{1950ApJ...112...72M} and more recently confirmed by \citet{1999A&AS..134..255K}, who listed it as a B-type star with a visual brightness of 10.7 mag. \citet{1989AJ.....98.1768C} tried to classify the object using optical multicolor photometry, which showed colors of a reddened early B--type star ($U-B = -0.09$ mag, $B-V = 0.89$ mag), but came up with a spectral type of F8. AS\,386 was also observed in the course of the Northern Sky Variability Survey \citep[NSVS, no filter optical photometry,][]{2004AJ....127.2436W} during 115 days in September--December 1999 and exhibited variations with an amplitude of 0.15 mag.

The IR-excess of AS\,386 was noticed by \citet{2005MNRAS.363.1111C} in the course of a positional cross-identification of the optical Tycho-2 \citep {2000A&A...355L..27H} and the IR MSX catalog \citep {2003yCat.5114....0E}. We independently identified this object in the NOMAD catalog \citep{z05} as an early-type star with a strong near-IR excess \citep{2017ASPC..508..229K} and considered it a candidate object with the B[e] phenomenon.

In this paper we report the results of low- and  high-resolution spectroscopy as well as multicolor optical and near-IR photometry of AS\,386 aimed at deriving its fundamental parameters and attempting to reveal its nature and evolutionary status. Multicolor photometry of stars around AS\,386 was used to constrain the interstellar extinction law in its direction and independently estimate the distance toward it.

\begin{figure}[!t]
\setlength{\unitlength}{1mm}
\resizebox{9.cm}{!}
{
\begin{picture}(70,70)(0,0)
\put (0,0){\includegraphics[width=7.cm]{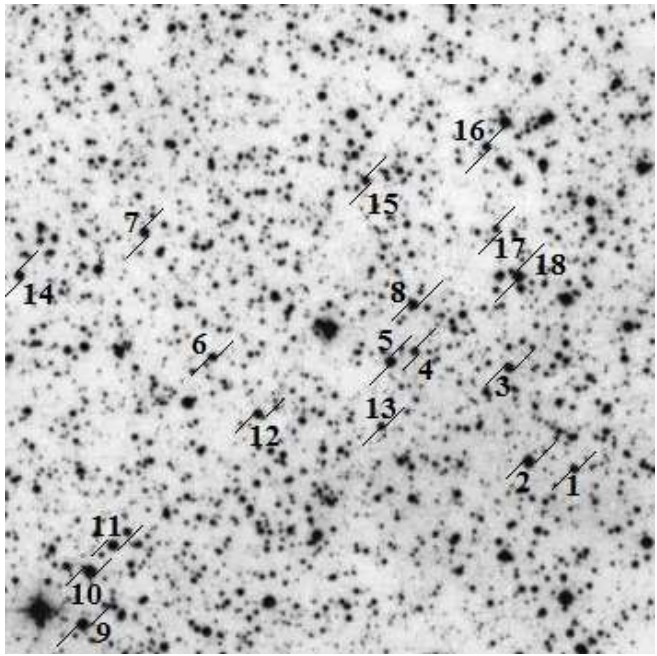}}
\end{picture}
}
\caption{Image of a $8\arcmin \times 8\arcmin$ field around AS\,386 in the $V$--band taken at TShAO. AS\,386 is the brightest star in the very center
of the image. $BVR_{\rm c}$ brightness of the numbered stars was measured to refine the interstellar extinction law in the object's direction.
North is at the top, East is at the left.}
\label{f1}
\end{figure}

\begin{figure}[!t]
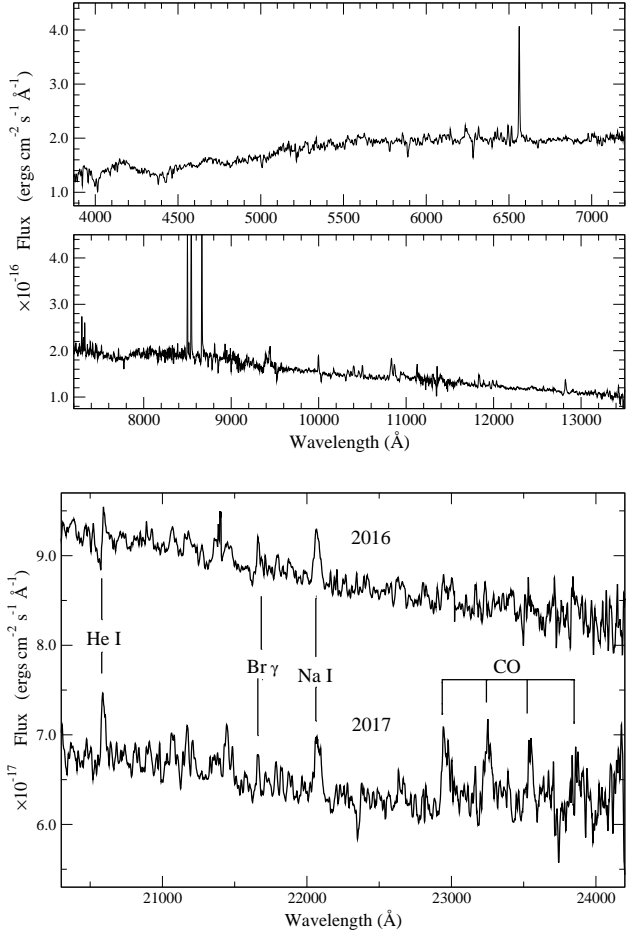

\setlength{\unitlength}{1mm}
\resizebox{9.cm}{!}
{
\begin{picture}(70,105)(0,0)
\put (0,0)       {\includegraphics[width=7.cm,bb = 0 40  705 565, clip=]{AS386fig2a.eps}}
\put (0,51.5)  {\includegraphics[width=7.cm,bb = 20 40  705 565, clip=]{AS386fig2b.eps}}

\end{picture}
}
\caption{Parts of absolutely calibrated low-resolution optical spectra of AS\,386 taken at the Lick Observatory. The spectra shown in the bottom panel
were taken at the opposite phases of the orbital cycle (0.53 in 2016 and 0.03 in 2017, see Sect.\,\ref{nir_sp}).}
\label{f2}
\end{figure}

\begin{table}[!h]
\caption[]{Optical Photometry of AS\,386}\label{t1}
\begin{center}
\begin{tabular}{ccrcr}
\hline\noalign{\smallskip}
JD    & Phase &V & $B-V$ &$V-R_{\rm c}$ \\
\noalign{\smallskip}\hline\noalign{\smallskip}
7486.433 &0.74  & 10.99  & 0.83   &  0.51\\
7627.290 &0.81  & 10.98  & 0.83   &  0.53\\
7666.136 &0.11  & 10.96  & 0.81   &  0.53\\
7673.106 &0.16  & 10.94  & 0.83   &  0.48\\
7693.048 &0.31  & 10.98  & 0.84   &  0.52\\
7694.072 &0.32  & 10.95  & 0.85   &  0.51\\
7695.042 &0.33  & 10.94  & 0.84   &  0.51\\
7702.081 &0.38  & 10.89  & 0.84   &  0.51\\
7703.030 &0.39  & 10.91  & 0.83   &  0.51\\
7733.494 &0.63  & 10.92  & 0.81   &  $-$ \\
7736.024 &0.64  & 10.92  & 0.85   &  0.51\\
7743.035 &0.69  & 10.96  & 0.81   &  0.53\\
7743.503 &0.70  & 10.95  & 0.80   &  $-$\\
7747.115 &0.72  & 10.87  & 0.84   &  0.51\\
7983.635 &0.53  & 10.96  & 0.83   & $-$ \\
8006.556 &0.70  & 10.92  & 0.83   & $-$ \\
\noalign{\smallskip}\hline
\end{tabular}
\end{center}
\begin{list}{}
\item The $BV(R_{\rm c})$ data obtained at TShAO and $BV$ data obtained at DSO in the Johnson-Cousins photometric system.
Column information: (1) -- Julian Date (JD$-$2450000), (2) -- orbital phase according to the RV solution (see text), (3--5) -- the photometric data.
Average measurement uncertainties do not exceed 0.02 mag in all the bands.
\end{list}
\end{table}

\begin{table}[t]
\caption[]{Near-IR Photometry of AS\,386}\label{t2}
\begin{center}
\begin{tabular}{cccccccc}
\hline\noalign{\smallskip}
JD    &Phase & $J$ &$\sigma J$& $H$ &$\sigma H$& $K$ &$\sigma K$\\
\noalign{\smallskip}\hline\noalign{\smallskip}
6051.544& 0.79& 8.52 &0.01 &7.76 &0.02 &6.87 &0.03\\
6223.462& 0.84& 8.51 &0.02 &7.75 &0.02 &6.87 &0.03\\
6565.994& 0.68& 8.54 &0.01 &7.75 &0.02 &6.90 &0.00\\
6679.362& 0.83& 8.43 &0.01 &7.64 &0.03 &6.82 &0.05\\
6886.091& 0.12& 8.54 &0.00 &7.71 &0.00 &6.84 &0.01\\
6912.399& 0.27& 8.59 &0.01 &7.76 &0.00 &6.90 &0.04\\
7016.371& 0.56& 8.57 &0.00 &7.79 &0.01 &6.96 &0.02\\
7174.426& 0.35& 8.59 &0.02 &7.83 &0.05 &6.96 &0.04\\
7180.441& 0.39& 8.59 &0.02 &7.83 &0.02 &6.99 &0.02\\
7190.509& 0.44& 8.60 &0.01 &7.90 &0.02 &7.04 &0.02\\
7203.545& 0.58& 8.57 &0.00 &7.81 &0.02 &6.96 &0.02\\
7256.443& 0.60& 8.57 &0.01 &7.83 &0.00 &6.97 &0.02\\
7358.277& 0.76& 8.49 &0.01 &7.71 &0.02 &6.86 &0.03\\
7560.495& 0.30& 8.55 &0.02 &7.78 &0.02 &6.95 &0.02\\
7654.393& 0.02& 8.51 &0.02 &7.69 &0.02 &6.79 &0.02\\
7658.430& 0.05& 8.54 &0.02 &7.70 &0.02 &6.80 &0.02\\
7888.519& 0.80& 8.50 &0.02 &7.67 &0.02 &$-$   & $-$\\
7891.524& 0.83& 8.52 &0.02 &7.68 &0.02 &$-$   & $-$\\
7896.069& 0.86& 8.52 &0.01 &7.70 &0.01 &6.83 &0.01\\
\noalign{\smallskip}\hline
\end{tabular}
\end{center}
\begin{list}{}
\item $JHK$ observations were obtained at Campo Imperatore. Column information: (1) -- Julian Date (JD$-$2450000),
(2) -- orbital phase according to the RV solution (see text), (3,4) -- $J$--band brightness and its 1 $\sigma$
uncertainty, (5,6) --  $H$--band brightness and its 1 $\sigma$ uncertainty, (7,8) -- $K$--band brightness and its 1 $\sigma$
uncertainty.
\end{list}
\end{table}

\begin{table*}[!htb]
\caption[]{Summary of the Spectroscopic Observations}\label{t3}
\begin{center}
\begin{tabular}{ccrllcrccrllcr}
\hline\noalign{\smallskip}
 Date       &JD          &Exp.  &\hspace{0.3cm}Range                &ID  & Ph.     & RV\hspace{0.27cm}             &Date       &JD          &Exp.                    &\hspace{0.3cm}Range                               &ID  & Ph.     & RV\hspace{0.27cm}             \\
                &             & sec\hspace{0.2cm}   & \hspace{0.4cm} \AA -- \AA  &     &          &km\,s$^{-1}$&                &                & sec   \hspace{0.2cm}& \hspace{0.4cm} \AA -- \AA  &           &          &km\,s$^{-1}$ \\
\noalign{\smallskip}\hline\noalign{\smallskip}
 1 & 2 & 3\hspace{0.25cm} &\hspace{0.4cm} 4 & 5 & 6 & 7\hspace{0.25cm} &  1 & 2 & 3\hspace{0.25cm} &\hspace{0.4cm} 4 & 5 & 6 & 7\hspace{0.25cm}  \\
\noalign{\smallskip}\hline\noalign{\smallskip}
11/05/09  &5140.621 & 1200      &3800--6800     & 1 & 0.876   &     4.6  &11/04/15  &7330.774  &1280        &3600--10500  & 3 & 0.554   & $-$16.1\\
06/22/10  &5369.971 & 20          &4600--25000  & 2  & 0.623   &  $-$      &12/02/15  &7358.703  &1280        &3600--10500  & 3 & 0.767   &      18.1\\
10/18/10  &5487.652  &4800      &3800--6800    & 1 & 0.519   & $-$23.9 &07/20/16  &7589.922 & 20         &4600--25000  & 2    & 0.532   & $-$\\
11/21/10  &5521.683  &540        &3600--10500  & 3 & 0.775   &      22.2 &09/09/16  &7640.733  &3600      &3600--7300   & 1  & 0.919   & $-$2.5\\
11/26/10  &5526.692  &540        &3600--10500  & 3 & 0.813   &      16.2 &09/22/16  &7653.746  &807        &3600--10500  & 3 & 0.014   & $-$36.0\\
12/16/10  &5546.732  &540        &3600--10500  & 3 & 0.965   & $-$16.9 &10/13/16  &7674.695  &1112      &3600--10500  & 3  & 0.174   & $-$76.9\\
07/06/12  &6114.921  &1800       &3600--10500  & 3 & 0.293   & $-$81.8 &10/15/16  &7676.675  &6000      &3600--7300   & 1    & 0.192   & $-$75.7\\
11/13/12  &6244.652  &4800       &3800--6800    & 1   & 0.285   & $-$84.0&10/19/16  &7680.609  &4800      &3600--7300   & 1    & 0.222   & $-$80.9\\
01/03/13  &6295.537  &2400       &3600--10500  & 4   & 0.673   &        7.4&11/18/16  &7710.619  &3600      &3600--7300   &  1     & 0.451   & $-$42.7\\
08/23/13  &6527.891  &2160       &3600--10500  & 3 & 0.439   & $-$56.0 &06/16/17  &7920.982 & 20           &4600--25000  & 2      & 0.053   & $-$ \\
09/14/13  &6549.902  &1320       &3600--10500  & 3 & 0.606   &         0.4 & 08/13/17  &7978.760  &1800        &3600--10500  & 3  & 0.489   & $-$40.8 \\
09/24/15  &7289.854  &1280       &3600--10500  & 3 & 0.242   & $-$90.2 & 08/15/17  &7980.962  &1800       &3600--10500  & 3  & 0.506   & $-$28.7 \\
10/03/15  &7298.706  &6000       &3800--7200    &  1  & 0.314   & $-$78.4& 09/08/17  &8004.752  &1800       &3600--10500  & 3  & 0.678   & 19.7 \\
10/08/15  &7303.725  &3600       &3800--7200    &  1  & 0.352   & $-$72.4& 11/13/17  &8070.666  &2000       &4000--10500  & 5    & 0.193   &  $-$89.0\\
10/30/15  &7325.838  &1280       &3600--10500  & 3 & 0.516   & $-$28.6& 11/28/17  &8085.569  &1500         &4000--10500  & 5   & 0.307   &  $-$86.9\\
\noalign{\smallskip}\hline
\end{tabular}
\end{center}
\begin{list}{}
\item Column information: (1) -- observing date (Month/Day/Year), (2) -- Julian Date (JD$-$2450000),
(3) -- total exposure time in seconds, (4) -- spectral range in \AA\,
(5) -- Observatory ID: 1 -- Observatorio Astron\'omico Nacional San Pedro Martir, 2 -- Lick Observatory,
3 -- Canada-France-Hawaii Telescope, 4 -- McDonald Observatory, 5 -- Apache Point Observatory,
(6) -- Orbital phase (see Sect.\,\ref{absorptions}), and (7) -- radial velocity (RV) of the B--type star.
\item Measurement uncertainties of the RV from column 7 are 1--2 km\,s$^{-1}$
\end{list}
\end{table*}

\begin{figure*}[t]
\setlength{\unitlength}{1mm}
\resizebox{9.cm}{!}
{
\begin{picture}(70,130)(0,0)
\put (5,0)   {\includegraphics[width=13.2cm, bb = 150 215 575 640, clip=]{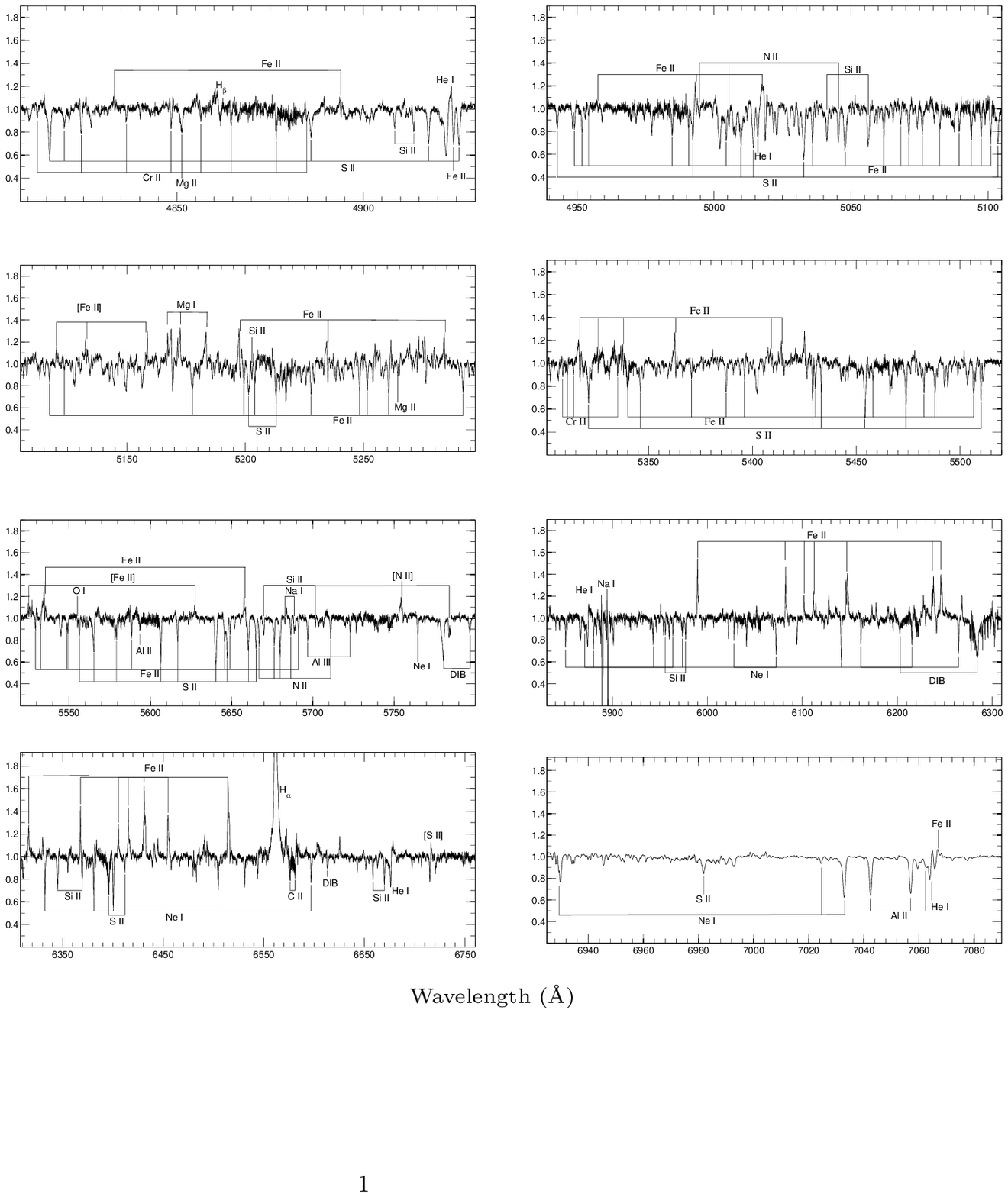}}
\end{picture}
}
\caption{Parts of the optical high-resolution spectrum of AS\,386 taken with CFHT on 12/02/15 except for the 6300--6760 \AA\ region, which is extracted from the CFHT spectrum taken on 09/24/15. The latter spectrum better shows a weak emission line of [S {\sc ii}] at 6717 \AA.} The intensity is normalized to the continuum, the wavelength scale is heliocentric.
\label{f3}
\end{figure*}

\section{Observations and Data reduction}\label{observations}

Figure\,\ref{f1} shows a $8\arcmin \times 8\arcmin$ field around AS\,386, which is the brightest star in the very center of the image.
Optical photometry of  the field in the Johnson--Cousins system was obtained on 12 nights at a 1\,m telescope of the Tien-Shan Observatory (TShAO) of the Fesenkov Astrophysical Institute of the National Academy of Sciences of Kazakhstan in April--December 2016. A $3056 \times 3056$ Apogee F9000 D9 CCD camera with 12 $\mu$m pixels and a 3--step cooling was used with a set of $BVR_{\rm c}$ filters. The TShAO data were reduced using
Maxim-DL\footnote{http://diffractionlimited.com/product/maxim-dl/} and converted into the standard Johnson-Cousins photometric system based on transformations
derived from observations of the open cluster NGC\,2169.

Additionally, four $UBV$ observations of the object's field were taken in December 2016 and August/September 2017 at a 40\,cm telescope of the Dark Sky Observatory
(DSO, near Boone, NC) in order to constrain the shortest-wavelength data point of its  spectral energy distribution (SED) and check the $U-B$ color reported by
\citet[][see Sect.\,\ref{intro}]{1989AJ.....98.1768C}.
These data were reduced with a standard pipeline provided by the Skynet network\footnote{http://skynet.unc.edu}.
The brightness of the object and field stars shown in Fig.\,\ref{f1} was measured using the IRAF\footnote{IRAF is distributed by the National Optical Astronomy Observatory,
which is operated by the Association of Universities for Research in Astronomy (AURA) under a cooperative agreement with the National Science Foundation.}
task {\it apphot/qphot}. Transformations between the instrumental and standard photometric system were determined by measuring the brightness of standard stars in
several open clusters (e.g., NGC\,2169) and using photometric data listed in the UCAC4 catalog \citep{2013AJ....145...44Z}. The derived $U-B$ color-index is
discussed in Sect.\,\ref{sed}. The optical photometric data from both TShAO and DSO are shown in Table\,\ref{t1}.

Near-IR $JHK$ photometry was taken with the 1.1\,m telescope AZT--24 at Campo Imperatore (Italy) in 2008--2017 with an imager/spectrometer SWIRCAM
\citep{2000SPIE.4008..748D}. The brightness of AS\,386 was measured by comparison with 2MASS data \citep{2003yCat.2246....0C} for several field stars. The results of the near-IR photometry are presented in Table\,\ref{t2}.

Twenty seven high-resolution optical spectra of AS\,386 with spectral resolving powers of $R = 18,000-65,000$ and an average signal-to-noise ratio of $\sim$100 were obtained between November 2009 and November 2017 (see Table\,\ref{t3}).
The observations were carried out using
the 2.1\,m telescope of the Observatorio Astron\'omico Nacional San Pedro Martir (OAN SPM, Baja California, M\'{e}xico) with a $R$=18,000 REOSC spectrograph,
the 2.7\,m Harlan J. Smith telescope of the McDonald Observatory (Texas, USA) with the $R$=60,000 Tull coud\'e spectrograph TS2 \citep{1995PASP..107..251T},
the 3.6\,m Canada-France-Hawaii Telescope (CFHT, Mauna Kea, HI, USA) with the $R$=65,000 ESPaDOnS spectropolarimeter,
and the 3.5\,m Astrophysical Research Consortium telescope at the Apache Point Observatory (APO, New Mexico, USA) with the $R$=31,500 ARCES spectrograph
\citep{2003SPIE.4841.1145W}.
The data obtained at OAN SPM, McDonald, and APO were reduced in a standard way using the $imred/echelle$ task of IRAF.
Observations obtained at CFHT were reduced with the Upena and Libre-ESpRIT software packages \citep{1997MNRAS.291..658D}.
Typical uncertainties of the wavelength calibration are $<$ 1 km\,s$^{-1}$ for the CFHT, McDonald, and APO data and $\sim$1 km\,s$^{-1}$ for OAN SPM.

Three low-resolution optical and near-IR spectra of AS\,386 were taken at the 3\,m telescope of the Lick Observatory (USA, 0.46--2.5 $\mu$m, R$\sim$700) in 2010, 2016, and 2017 (see Table\,\ref{t3}) with The Aerospace Corporation's Visible and Near Infrared Imaging Spectrograph \citep[VNIRIS,][]{1999AJ....118..666R}. The spectra were absolutely calibrated using solar type standard stars (16 Cyg A, HIP\,109281, and HIP\,98894).
Parts of the optical low- and high-resolution spectra of AS\,386 are shown in Fig.\,\ref{f2} (top panels) and Fig.\,\ref{f3}, respectively.

\begin{table}[t]
\caption[] {Lines in the spectrum of AS\,386}\label{t4}
\begin{center}
\begin{tabular}{lrcccl}
\hline\noalign{\smallskip}
Line ID      & Mult. &$\lambda_{\rm lab}$ & EW & I/I$_{\rm c}$ & Comment\\
\noalign{\smallskip}\hline\noalign{\smallskip}
He {\sc i}	&22	   &3819.61       &0.70  &0.55 &      \\
He {\sc i}	&62	   &3833.57       &0.33  &0.71 &      \\
H  {\sc i}	&2	   &3835.39       &0.30  &0.66 &  H9\\
S  {\sc ii}	&22	   &3845.21       &0.08  &0.80 &      \\
Mg {\sc ii}	&5	   &3848.24       &0.08  &0.80 &      \\
Si {\sc ii}	&1	   &3853.66       &0.28  &0.61 &      \\
Si {\sc ii}	&1	   &3856.02       &0.36  &0.49 &      \\
Si {\sc ii}	&1	   &3862.59       &0.35  &0.45 &      \\
He {\sc i}	&20	   &3867.48       &0.24  &0.61 &      \\
....             &&&&&\\
\noalign{\smallskip}\hline
\end{tabular}
\end{center}
\begin{list}{}
\item Column information: (1) -- element and ionization state, (2) -- multiplet number (for the lines identified in \citet{1993BICDS..43....7C},
(3) laboratory line position in \AA, (4) -- line EW in \AA, (5) -- peak intensity of the line with respect to the local continuum, and (6) -- comment on the line
appearance in the spectrum or its ID.
\item Line properties were measured in the CFHT spectrum taken on 07/06/12.
\tablecomments{\hsize 3in
 Table\,\ref{t4} is published in its entirety in the electronic edition of the
Astrophysical Journal. A portion is shown here for guidance regarding its form and content.}
\end{list}
\end{table}

\begin{figure}[!h]
\setlength{\unitlength}{1mm}
\resizebox{8.cm}{!}{
\begin{picture}(70,55)(0,0)
\put (0,5)       {{\includegraphics[width=7.3cm, clip=]{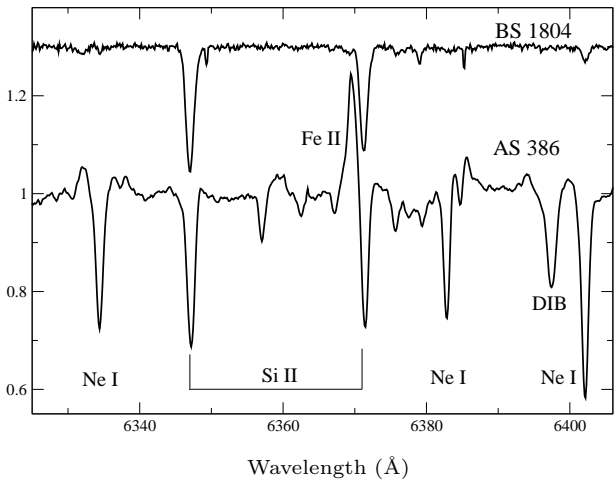}}}
\put (29, 0)   {\scriptsize Wavelength (\AA)}
\end{picture}}
\caption{{Comparison of the spectrum of AS\,386 with that of BS\,1804 (B9 {\sc i}b). The spectra are continuum
normalized and shifted with respect to each other along the vertical axis. The spectrum of BS\,1804 was taken at OAN SPM. The spectrum of AS\,386
was averaged from several spectra that were shifted to the same (zero) heliocentric RV.
Intensities and wavelengths are shown in the same units as in Fig.\,\ref{f3}.
}
\label{f4}}
\end{figure}

\begin{figure}[t]
\setlength{\unitlength}{1mm}
\resizebox{9.cm}{!}{
\begin{picture}(70,100)(0,0)
\put (1,73.5)   {{\includegraphics[width=6.5cm, bb= 30 50 720 325, clip= ]{AS386fig5a.eps}} }
\put (0,46.5  ) {{\includegraphics[width=6.5cm, bb= 20 44 705 330, clip= ]{AS386fig5b.eps}}}
\put (0,23.75) {{\includegraphics[width=6.5cm, bb= 20 44 705 330, clip= ]{AS386fig5c.eps}}}
\put (2,-7)       {{\includegraphics[width=6.8cm, bb=1  270 550 510, clip= ]{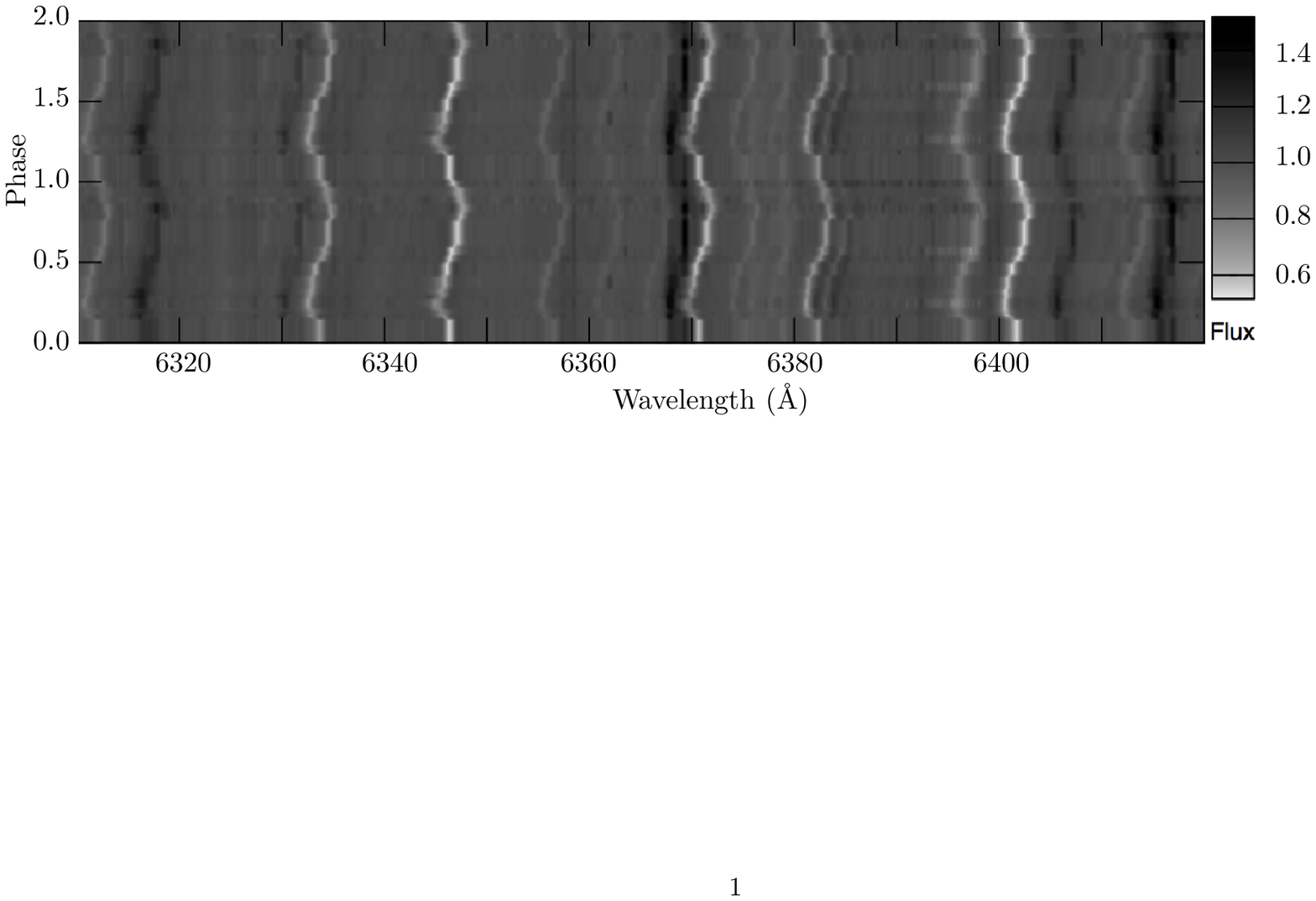}}}
\end{picture}}
\caption{{Top panel: Lomb-Scargle periodogram calculated for the RV data derived by cross-correlation. The horizontal dashed line show the peak significance level.
Middle panels: top -- Peak intensity ratios (V/R) of the Fe {\sc ii} 6432 \AA\ (filled circles) and [Ca~{\sc ii}]~7291 \AA\ (open squares) folded with the orbital period;
bottom -- RV curve derived from our spectra of AS\,386 folded with the orbital period. The dashed line shows the systemic RV,
and the solid line represents the best fit to the data with the orbital parameters described in text.
Bottom panel: Trailed spectra of the 6310--6420~\AA\ region folded with the orbital period. Intensity (relative flux) scale is shown on the right side, and phase is
shown on the left side.
}
\label{f5}}
\end{figure}

\begin{figure}[t]
\setlength{\unitlength}{1mm}
\resizebox{9.cm}{!}{
\begin{picture}(70,45)(0,0)
\put (-2,0) {\includegraphics[width=7.0cm,  clip=]{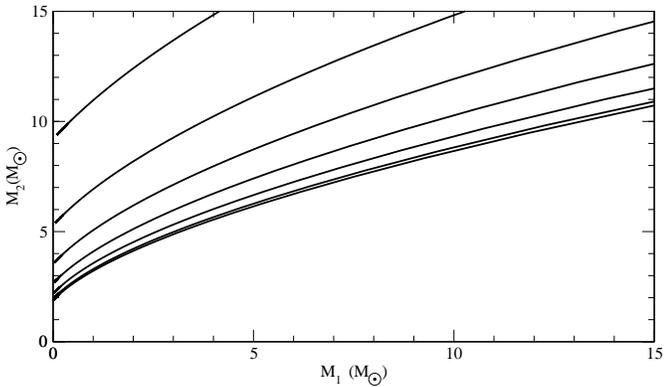}}
\end{picture}}

\caption{{M$_2$ versus M$_1$ (B-type component) relationship derived from the mass function. The lines are shown
for orbital inclination angles separated by 9$\degr$. The lowest line is shown for the 90$\degr$ inclination of the orbital axis with respect to the line of sight.}
\label{f6}}
\end{figure}

\begin{figure}[!h]
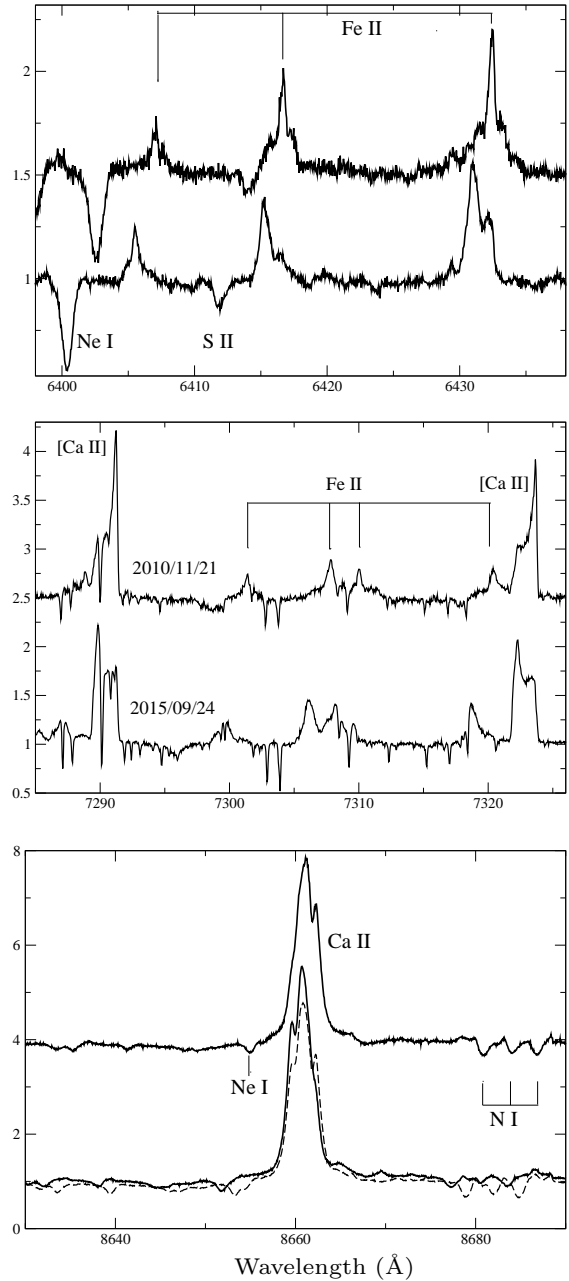

\setlength{\unitlength}{1mm}
\resizebox{9.cm}{!}
{
\begin{picture}(70,138)(0,0)
\put (2, 95)  {\includegraphics[width=6.0cm, clip=]{AS386fig7a.eps}}
\put (2, 50 ) {\includegraphics[width=6.0cm, clip=]{AS386fig7b.eps}}
\put (2, 3 )   {\includegraphics[width=6.0cm, clip=]{AS386fig7c.eps}}
\put (26, 0)   {\scriptsize Wavelength (\AA)}
\end{picture}
}
\caption{Parts of high-resolution CFHT spectra of AS 386 at different orbital phases. Spectra taken on the same dates at the opposite phases are shown by the solid
lines in each panel. The dashed line in the bottom panel shows the CFHT spectrum taken on August 13, 2017, when the star's RV was near the systemic velocity.
Intensities and wavelengths are shown in the same units as in Fig.\,\ref{f3}.
}
\label{f7}
\end{figure}

\section{Data Analysis}\label{analysis}

\subsection{Spectroscopic results}\label{sp_results}

The optical spectrum exhibits numerous emission and absorption lines including diffuse interstellar bands (DIBs). They were identified using a catalog by \citet{1993BICDS..43....7C} and databases by  P. van Hoof\footnote{http://www.pa.uky.edu/$\sim$peter/newpage/} and at the National Institute of Standards and Technology
\footnote{Kramida, A., Ralchenko, Yu., Reader, J., and NIST ASD Team (2015). NIST Atomic Spectra Database (ver. 5.3), available at http://physics.nist.gov/asd}.
Overall, the absorption line content indicates a late B spectral type and a moderately high luminosity. A large part of the high-resolution spectrum is
shown in Fig.\,\ref{f3}. A list of $\sim$450 identified lines in the optical spectrum is presented in Table\,\ref{t4}. The line content in the optical and near-IR regions is
discussed separately in Sect.\,\ref{absorptions}--\ref{nir_sp}.

\subsubsection{Absorption lines} \label{absorptions}

The main features of the absorption-line spectrum in the optical region are lines of neutral helium and singly ionized metals (e.g., Si {\sc ii}, N {\sc ii}, Ne {\sc i}, S {\sc ii})
with a smaller number of doubly ionized metals (e.g., Si {\sc iii}, Al {\sc iii}). Hydrogen lines of the Balmer series are partially or fully in emission. Such a content is typical
for B-type stars. Relatively strong lines of Si {\sc ii} 6347/6371 \AA\ (average EW $\sim$0.4 \AA), which are a few times weaker in spectra of B-type dwarfs, indicate a luminosity above the main sequence \citep{1974ApJ...187..261R,2013msao.confE.169M}. Therefore, we compare the object's spectrum with those of B-type supergiants
in order to determine its fundamental parameters.

Nearly the same EWs of the He {\sc i} 4471 \AA\ and the Mg {\sc ii} 4481 \AA\ (their EW ratio is 1.04$\pm$0.05), moderately weak He {\sc i} lines, the absence of Si {\sc iv}
lines, weak lines of the Si {\sc iii} 4552/4568/4575~\AA\ triplet (EW $\sim$0.1 \AA) and virtually absent lines of the Si {\sc iii} 4813/4819/4828~\AA\ triplet, a small ratio of
the Si {\sc iii} 4568/Si {\sc ii} 4128~\AA\ line EWs ($\sim$0.2--0.3), and weak C {\sc ii} 6578/6582~\AA\ lines (EW $\sim$0.1 \AA) suggest an effective temperature of
T$_{\rm eff} = 11000\pm500$ K \citep[e.g.,][]{1993A&AS...97..559L,1999A&A...349..553M,2008A&A...478..823M,2013msao.confE.169M}.
Also, comparison with normal stars shows that lines of such elements as Ne {\sc i} (e.g., Multiplet 1, 5882/5945/5976/6143/6217/6334/6402/7032~\AA, see Fig.\,\ref{f4})
and Al {\sc ii} (e.g., Multiplet 3, 7042/7057/7064~\AA) are noticeably stronger in the spectrum of AS\,386.
This may indicate an overabundance of the mentioned metals.

The presence of the near-IR oxygen triplet at 7770--7775 \AA\ as well as the Si {\sc ii} lines at 6347 and 6371 \AA\ (all of moderate strengths in absorption, see, e.g., Fig.\,\ref{f4}) indicate a luminosity above the main sequence. In particular, luminosity calibrations of the oxygen triplet EW \citep[e.g.,][]{2003RMxAA..39....3A} imply a visual absolute magnitude of M$_{V} = -3.6\pm0.3$ mag. With a bolometric correction of BC = $-0.5\pm0.1$ mag \citep{1997IAUS..189P..50M}, the luminosity comes to $\log$ L/L$_{\odot} = 3.5\pm0.1$.
Therefore, a distance to AS\,386 estimated from the spectroscopic parallax is $D = 2.3\pm0.3$ kpc. This result is refined below (Sect.\,\ref{extinction}) based on a study of
the interstellar extinction law in the object's direction.

With the T$_{\rm eff}$ derived above, AS\,386 can be classified as a B8/9 {\sc i}b star \citep[e.g.,][]{2008A&A...478..823M} with an intrinsic color-index
$(B-V)_{0} = -0.02\pm0.01$ mag \citep{1977mspp.book.....S,1994MNRAS.270..229W}. Averaging our photometric data (Table\,\ref{t1}), we derive $B-V = 0.83\pm0.02$
mag and $E(B-V) = 0.85\pm0.03$ mag.

Analyzing the absorption lines, we found that positions of all of them (excluding those of interstellar features) are noticeably variable with the same amplitude and in the same phase (see Fig.\,\ref{f5}). For accurate RV measurements we used the cross-correlation method implemented in the IRAF package {\it rvsao} on several wavelength regions well populated with absorption lines. The following regions were used for this purpose: 5560--5715 \AA\ (containing 20 lines), 6310--6420 \AA\ (10 lines), 6620--6720 \AA\ (16 lines), 7020--7060 \AA\ (5 lines), and 7420--7490 \AA\ (6 lines, CFHT data only). Spectral regions that contain clustered emission lines and telluric features were avoided. The CFHT spectrum taken on 09/14/13 was used as a template. RVs of 40 spectral lines were measured in this spectrum by fitting their profiles to a Gaussian.
The measurement results are shown in Table\,\ref{t3}. No lines moving in anti-phase have been detected in the entire optical range. Therefore, AS\,386 is a single-lined
binary system.

The Lomb-Scargle periodogram analysis \citep{1982ApJ...263..835S} was used to search for periodicity in the RV data. The strongest peak in the power spectrum corresponds to a period of 131.3 days (Fig.\,\ref{f5}, upper panel). The set of RVs was then fitted to a theoretical RV curve expected from a binary system. The best fit was found for a circular orbit with the following elements: orbital period P$_{\rm orb} = 131.274\pm0.090$ days, zero-phase epoch (the B-type component is in a superior conjunction) JD$_{0}$ = JD$2455025.6\pm1.5$, semi-amplitude of the RV curve $K_{1} = 51.7\pm3.0$ km\,s$^{-1}$, and systemic velocity $\gamma = -31.8\pm2.6$ km\,s$^{-1}$.  The RV curve folded with the orbital period is shown in the middle panel of Fig.\,\ref{f5}, and trailed spectra of one of the regions used for cross-correlation is shown in the bottom panel of this~Figure. The resulting mass function, $f(m) = 1.9\pm0.3$ M$_{\odot}$, implies that the undetected component is relatively massive (see Fig.\,\ref{f6}).

\begin{figure*}[t]
\setlength{\unitlength}{1mm}
\resizebox{9.cm}{!}{
\begin{picture}(70,76)(0,0)
\put (-2,0){{\includegraphics[width=14.7cm, bb = -260 124 850 800, clip=]{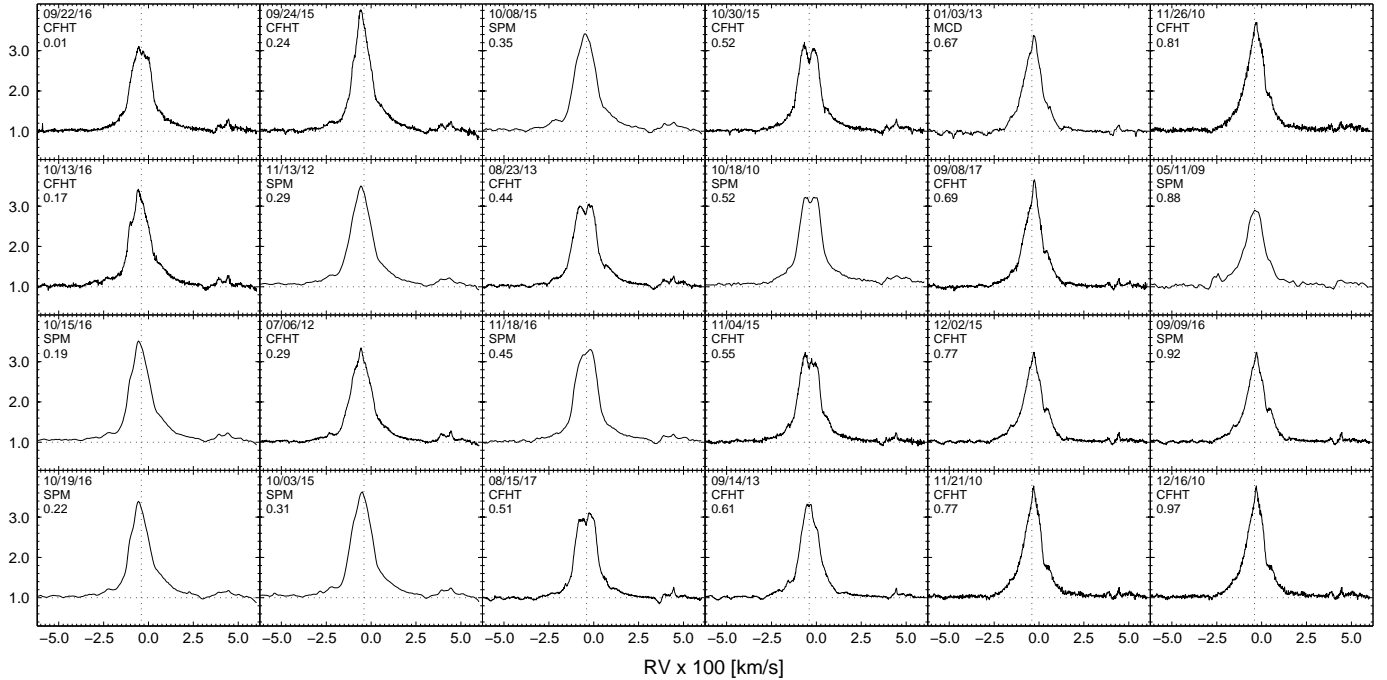}}}
\end{picture}}
\caption{Orbital variations of the H$\alpha$ line in the spectrum of AS\,386. Horizontal dotted lines show the continuum level across the line profiles, and vertical dotted lines show the systemic RV. Observatory names and orbital phases are shown along with the observing dates for each spectrum.
}
\label{f8}
\end{figure*}

\subsubsection{Emission lines}\label{emission}

Permitted emission lines in the spectrum of AS\,386 are represented by hydrogen (Balmer series), numerous Fe {\sc ii} (see Fig.\,\ref{f3}), and Ca {\sc ii} (IR triplet) lines.
The H$\alpha$ line is relatively weak with an average EW = 7.2$\pm$0.6 \AA\ and a profile that shows slight phase-locked changes (see Fig.\,\ref{f7} and \ref{f8}).
Hydrogen lines of the Paschen series were not detected in our high-resolution optical spectra, but a few weak lines of the lower members of the series were detected in the
near-IR region.

Forbidden lines are less numerous ([Fe {\sc ii}], [N {\sc ii}]), but some of them are rather strong ([Ca {\sc ii}] 7291.43, 7323.89~\AA, see middle panel of Fig.\,\ref{f7}).
The latter two lines have only been found in luminous (non-dwarf) objects with the B[e] phenomenon and are supposed to be formed in dense circumstellar disks \citep{2016MNRAS.456.1424A}. The forbidden lines are stationary with the blue component at a RV of $-68\pm1$ km\,s$^{-1}$ and the red component at a RV of
$+10\pm2$ km\,s$^{-1}$. The components are positioned nearly symmetrically about the systemic velocity ($-31.5$~km\,s$^{-1}$).

Most of the emission lines are double-peaked with a variable peak intensity ratio. Permitted lines (including the Balmer lines) move in phase with the
atmospheric lines of the hot component but with a smaller amplitude. Also, both permitted and forbidden lines show orbital modulation of the peak intensity ratios (V/R,
see Fig.\,\ref{f5}, second panel from top). The blue-shifted peak is stronger than the red-shifted one when the B-type component moves toward the observer. The situation reverses when it moves away from us.

\subsubsection{Near-IR spectrum}\label{nir_sp}

The three low-resolution spectra taken at Lick (see Sect.\,\ref{observations} and Fig.\,\ref{f2}) are all very similar to one another.
They contain a number of weak emission features which mostly include H {\sc i} and He {\sc i} lines.
We also identified a Na {\sc i} (2.2065 $\mu$m) line and the CO first overtone band heads (2.29--2.41 $\mu$m) in emission. The latter are typical in B[e] stars.

The 2017 Lick spectrum was taken at the phase 0.05, when the B-type component was located in a superior conjunction. The other two spectra were taken in
almost the opposite system configuration.  The He {\sc i} lines (1.003, 1.083, 2.058 $\mu$m) in the 2017 spectrum show blue-shifted absorption components,
which are not observed in 2010 and 2016. At the same time, some hydrogen lines (e.g., P5 at 1.282 $\mu$m) exhibit a blue-shifted absorption in all the
spectra. This is probably a sign of the B-type component wind. The CO lines are the strongest in the 2016 spectrum (phase 0.53).

\begin{figure}[t]
\setlength{\unitlength}{1mm}
\resizebox{9.cm}{!}{
\begin{picture}(70,52)(0,0)
\put (0,0) {{\includegraphics[width=6.8cm, bb = 25 55 705 560, clip= ]{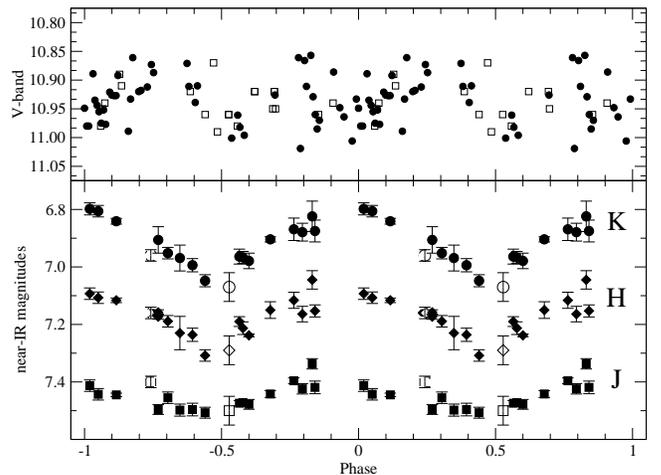}}}
\end{picture}}
\caption{{Photometric data folded with the system orbital period. Upper panel shows the optical light curve. NSVS data are shifted by +0.21 mag and
shown by filled circles. TShAO and DSO $V$--band data are shown by open squares.
Lower panel shows the near-IR light curves. Circles show the $K$--band data, diamonds show the $H$--band data (shifted by $-$0.6 mag),
and squares show $J$--band data (shifted by $-$1.1 mag). Filled symbols show the data from Table\,\ref{t4}, open symbols show
integrated fluxes derived from absolutely calibrated Lick spectra, and dashed symbols show the 2MASS photometry.
}
\label{f9}}
\end{figure}

\begin{figure}[!h]
\setlength{\unitlength}{1mm}
\resizebox{9.cm}{!}
{
\begin{picture}(70,47)(0,0)
\put (-2,0) {\includegraphics[width=7.cm, bb = 130 480 550 755, clip=]{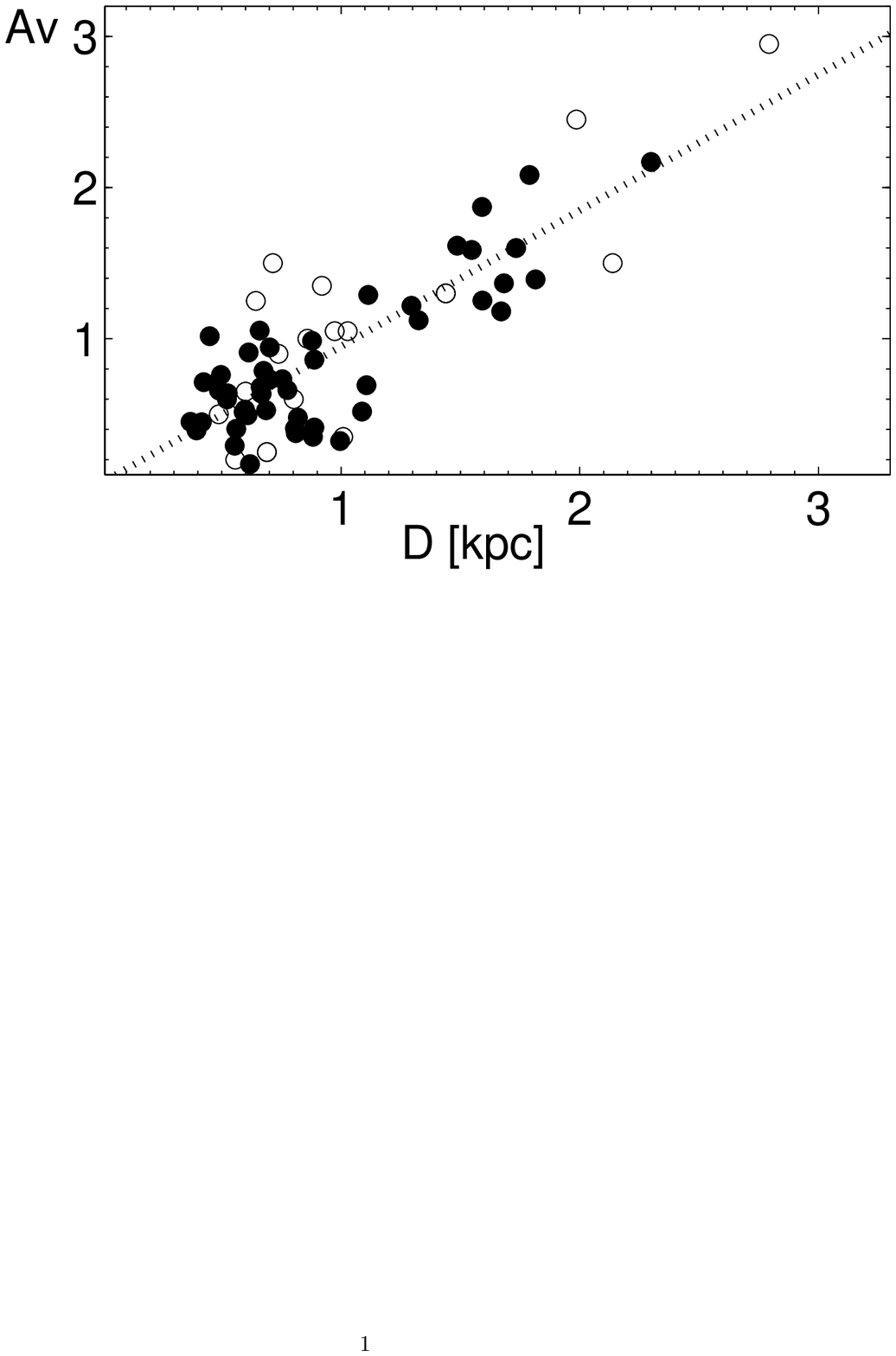}}
\end{picture}
}
\caption{{Interstellar extinction in the direction of AS\,386. Open circles show stars in the $8\arcmin \times 8\arcmin$ field around the object.
Filled circles show stars in a wider area of a $\sim 1\degr$ radius.}
\label{f10}}
\end{figure}

\begin{figure}[t]
\setlength{\unitlength}{1mm}
\resizebox{9.cm}{!}{
\begin{picture}(70,50)(0,0)
\put (0,0) {{\includegraphics[width=6.8cm, bb = 30 40 705 560, clip= ]{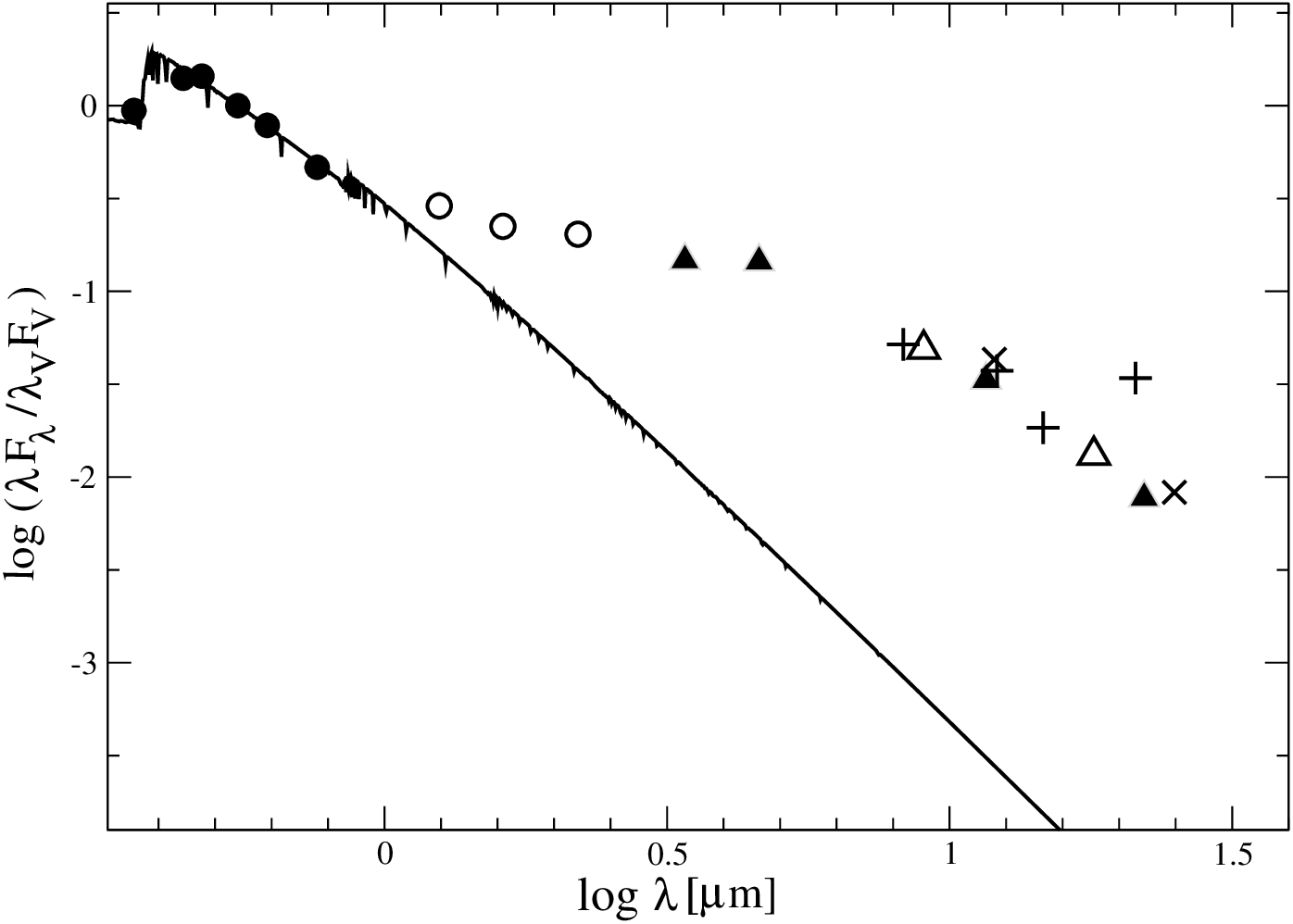}}}
\end{picture}}
\caption{Spectral energy distribution of AS\,386 corrected for an interstellar extinction of $E(B-V) = 0.85$ mag. Symbols: filled circles -- averaged ground-based optical photometry, open circles -- averaged $JHK$ photometry, filled triangles -- WISE data \citep{2012yCat.2311....0C}, open triangles -- AKARI data {\citep{2010A&A...514A...1I}, pluses -- MSX data, and crosses -- IRAS data \citep{1988iras....7.....H}.
A model atmosphere for T$_{\rm eff} = 11000$ K and $\log$ g = 3.0 that represents the hot component is shown by the solid line
 \citep{1998HiA....11..646K}. The fluxes are normalized to that in the $V$ band, the wavelengths are shown in microns.}
\label{f11}}
\end{figure}

\begin{figure}[t]
\setlength{\unitlength}{1mm}
\resizebox{9.cm}{!}{
\begin{picture}(70,54)(0,0)
\put (-2,2) {\includegraphics[width=6.95cm,  clip=]{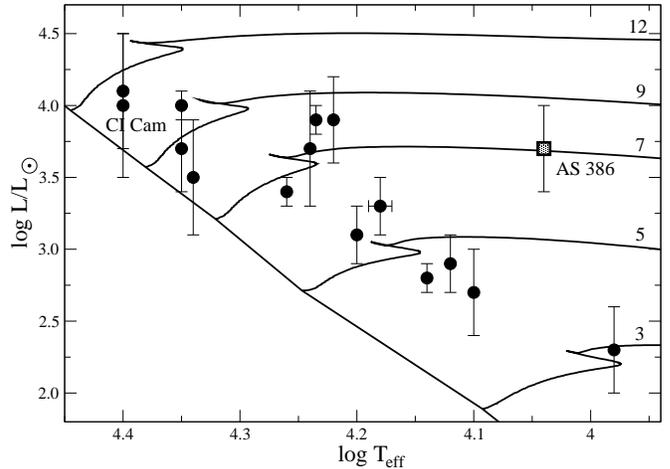}}
\end{picture}}
\caption{{Hertzsprung-Russell diagram with positions of FS\,CMa objects with known fundamental parameters. Evolutionary tracks for single rotating stars are taken from \citep{2012A&A...537A.146E} and shown by solid lines along with the main-sequence location. Numbers by the tracks indicate initial masses in solar units.
\label{f12}}}
\end{figure}

\subsection{Brightness variations}\label{brightness}

The optical brightness of AS\,386 is variable with an amplitude of $\Delta V \simeq 0.15$ mag and a standard deviation from the average brightness of 0.05 mag.
This result is derived from comparison of both available data sets (NSVS and our).
The NSVS data were shifted to take into account the passband of the survey and match the average level of our TShAO and DSO data (see Fig.\,\ref{f9}).
Periodogram analysis of the combined data set shows the presence of a broad but insignificant peak at a period of 61.9 days, which is roughly half the system orbital period. These variations show a fainter state of the object at phases near 0.0 and 0.5 and a brighter state at phases near 0.25 and 0.75. No convincing variations
of the $B-V$ color-index has been found.

If real, such a variability can be explained by eclipses in the system. However, the NSVS set of 45 data points barely covers only one orbital cycle. Our data set is even smaller and does not provide a significant phase coverage. Also, there is no spectroscopic confirmation of the secondary component contribution to the system optical flux in both the
optical and near-IR region. Therefore, the observed optical variations should probably be attributed to changing conditions in the circumstellar environments. More optical photometry data are needed to further conclude on the significance of this cycle.

Our near-IR photometric data set that contain 20 points and cover 14 orbital cycles clearly shows orbitally modulated brightness variations. The largest amplitude is detected in the $K$ band ($\sim$0.30 mag, see bottom panel of Fig.\,\ref{f9}). An amplitude is $\sim$0.25~mag in the $H$ band and $\sim$0.1~mag in the $J$ band. The IR light curves show a maximum near a phase of 0.0 and a minimum near a phase of 0.5. The timespan of the minimum is too long ($\sim$0.2--0.3 of the orbital period length)
to be attributed to an eclipse.

\subsection{Interstellar extinction law in the direction of AS\,386}\label{extinction}

We constructed the interstellar extinction law in the object's direction (see Fig.\,\ref{f10}) to independently estimate the distance to it. To do that, we collected published photometric data and MK types for hot stars in a 1$\degr$--size region around the object. Additionally, we measured the $BVR_{\rm c}$ brightness of 18 stars in the $8\arcmin \times 8\arcmin$ field shown in Fig.\,\ref{f1}. We added 2MASS $JHK$ data to our optical photometry and determined $T_{\rm eff}$ of these stars by comparison with theoretical SEDs from \citet{1998HiA....11..646K}.

The observed SEDs were de-reddened using a standard Galactic interstellar extinction law \citep{1979ARA&A..17...73S}. Distances to the stars were determined using a MK type luminosity calibration from \citet{1981Ap&SS..80..353S}.
A distance of $D = 2.5\pm0.3$ kpc to AS\,386 was found from a linear fit to the data shown in Fig.\,\ref{f10} assuming that the observed optical reddening, $E(B-V) = 0.85\pm0.03$ mag (Sect.\,\ref{absorptions}), which implies A$_V = 2.7\pm0.1$ mag that is solely due to the interstellar medium. The latter assumption is justified by the analysis of the object's spectrum in Sect.\,\ref{absorptions}.

\subsection{Spectral Energy Distribution}\label{sed}

The SED of AS\,386 was composed from various sources, which included both our data and fluxes from IR all-sky surveys.
Our TShAO photometry turned out to be very close to the data listed for the object in the UCAC4 catalog, which contains 5-band ($BVgri$) photometry from the AAVSO Photometric All-Sky Survey. Additionally, we took images in the Johnson $U$--band to extend the wavelength coverage and check the unusually negative $U-B$ color-index taken on 1988 June 24 \citep[][see Sect.\,\ref{intro}]{1989AJ.....98.1768C}.

The $U-B$ color-index of AS\,386 was measured by comparison to that of star \#18 marked on the map shown in Fig.\,\ref{f1}. Its SED was composed
of the UCAC4, 2MASS, and WISE catalog data. Then it was fitted to a set of theoretical spectra \citep{1998HiA....11..646K} allowing for the interstellar extinction to be present.
The best fit for the SED of star \#18 was found for T${\rm eff} = 8500\pm500$ K, $\log$ g = 4.0, and A$_V = 0.80\pm0.05$ mag ($E(B-V) = 0.22$ mag).
This result implies that the intrinsic color-index $(U-B)_0 = 0.08$ mag \citep{1998A&A...333..231B} and that the observed one should be $(U-B) = 0.24$ mag (from a typical interstellar color-excess ratio of $E(U-B)/E(B-V) = 0.72$). Our measurements of the DSO data (see Sect.\,\ref{observations}) show that the color-index difference between
AS\,386 and star \# 18 is $\Delta (U-B) = 0.08\pm0.01$ mag. Therefore, our result for the observed color-index of AS\,386 is $(U-B) = 0.32\pm0.05$ mag. The uncertainty is mostly determined by the SED of star \# 18 fitting errors. The derived $(U-B)$ along with the average $(B-V) = 0.83\pm0.02$ mag from our TShAO data and the mentioned above color-excess ratio leads to the intrinsic color-index $(U-B)_{0} = -0.30\pm0.07$ mag.
The average color-index $(V-R_{\rm c}) = 0.51\pm0.02$ mag corresponds to an intrinsic value of $(V-R_{\rm c})_0 = 0.02\pm0.05$ mag
\citep[using $E(V-R_{\rm c})/E(B-V) = 0.56$,][]{1986ApJS...60..577T}. Both intrinsic color-indices are consistent with the fundamental parameters of the B-type component of AS\,386 found from the high-resolution spectroscopy (see Fig.\,\ref{f11}).

\subsection{Distance and Fundamental parameters}\label{fund_dist}

The distance determined in Sect.\,\ref{extinction} ($D = 2.5\pm0.3$ kpc) is very similar to that found from the spectral features ($D = 2.3\pm0.3$ kpc).
Averaging the two, the final distance toward AS\,386 comes to $D = 2.4\pm0.3$ kpc. Note that the parallax of AS\,386 listed in the GAIA first
data release turned out to be negative \citep[$\pi = -0.47\pm0.33$ mas,][]{2016A&A...595A...2G}. This measurement was probably affected by the star's orbital motion and the large distance derived here.

This distance estimate is qualitatively supported by the strengths of the DIBs (e.g., 5780, 5797, 6613 \AA, see Table\,\ref{t4}), and the complicated structure of the Na {\sc i} D--lines at 5889 and 5895 \AA. Both type of these features indicate a moderate interstellar reddening, although their strengths do not give a very precise estimate for it. Nevertheless, a frequently used calibration by \citet{1993ApJ...407..142H} for the 5780 and 5797 \AA\ DIB EWs  ($0.45\pm0.05$ \AA\ and $0.15\pm0.03$ \AA) returns a reddening of $E(B-V) = 0.9\pm0.2$ mag, which is in a good agreement with that found above from the optical photometry. No features of the secondary component detected in the optical spectrum of AS\,386 (see above), the absence of eclipses in the optical photometry (see Sect.\,\ref{brightness}), and the weak emission lines justify our assumption of a negligible contribution from the secondary component and the circumstellar material to the observed reddening.

Our result for the distance of AS\,386 implies a luminosity of $\log$ L/L$_{\odot} = 3.7\pm0.3$. The object's location in the Hertzsprung-Russell diagram is shown
in Fig.\,\ref{f12}.
It gives a current evolutionary mass of 7$\pm$1 M$_{\odot}$, a radius to $R = 19\pm5 R_{\odot}$, and a surface gravity of $\log$ g = 2.7$\pm$0.3.
A projected rotational velocity of $v \sin\,i = 25\pm3$ km\,s$^{-1}$ determined from the profiles of the He {\sc i}, Si {\sc ii}, and Ne {\sc i} lines using the Fourier transform technique implies a slow rotation of the star (see Fig.\,\ref{f4} and Sect.\,\ref{dust_temp} for the rotational axis tilt angle) and is consistent with the position of AS\,386 beyond the main sequence.

\subsection{Secondary component}\label{secondary}

The photometric behavior in the visual spectral region (see Sect.\,\ref{brightness}) along with the weak variations of the spectrum suggest that the B-type component continuum itself is not the main source of the near-IR brightness variations. Therefore, they could be due to either 1)~a variable contribution from a secondary component or 2)~a variable illumination of the inner rim of the dusty disk, which surrounds the binary system.

The former hypothesis implies that the secondary component is much fainter than the B-type component in the optical region,
and its IR flux varies due either to illumination of a part of its surface visible to the observer or to its own variability. However, the main problem with the
secondary component is a high mass inferred from the system mass function. The B-type component's position on HRD (see Fig.\,\ref{f12}) suggests its current mass to be
$7\pm1$ M$_{\odot}$.
Therefore, the secondary component should have nearly the same mass, even if the system orbital plane is viewed edge-on (see Fig.\,\ref{f6}).
Its bolometric luminosity should also be comparable with that of the B-type component. Since we detected no signs of an active mass transfer, both
components should be in equilibrium. The non-detection of spectral features of the secondary component in the optical and near-IR spectral range might imply that it is
either much hotter or much cooler than the B-type component.

In the cooler case, the secondary component would dominate the IR radiation from the system. However, its T$_{\rm eff}$ should not be below 3000 K,
because such cool stars show a significant continuum distortion near $\lambda = 2 \mu$m \citep[e.g.,][]{1992A&AS...96..593L} that is not observed in our spectra.
If its T$_{\rm eff}$ is between 3000 K and 4000 K, we would see a periodic increase of its near-IR flux near a phase of 0.5 due to illumination of its surface by the B-type
component that is the opposite of the observed near-IR light curve (see Fig.\,\ref{f9}). Even a higher T$_{\rm eff}$ of the secondary component would result in an increase of
its contribution to the system optical radiation and lead to either regular brightness variations (eclipses) and/or the presence of its lines in the spectrum of the entire system.
An example of the latter case is the FS\,CMa type binary MWC\,728, where a less massive G8-type secondary component is a few times fainter than $\sim$ 5 M$_{\odot}$
B5-type primary. Spectral lines of both components are easily detected throughout the optical region \citep{2015ApJ...809..129M}.
However, we observe neither of the described effects.

\begin{figure*}[t]
\setlength{\unitlength}{1mm}
\resizebox{9.cm}{!}{
\begin{picture}(70,95)(0,0)
\put (-5,0) {{\includegraphics[width=14.1cm, bb=10 335 585 920, clip=]{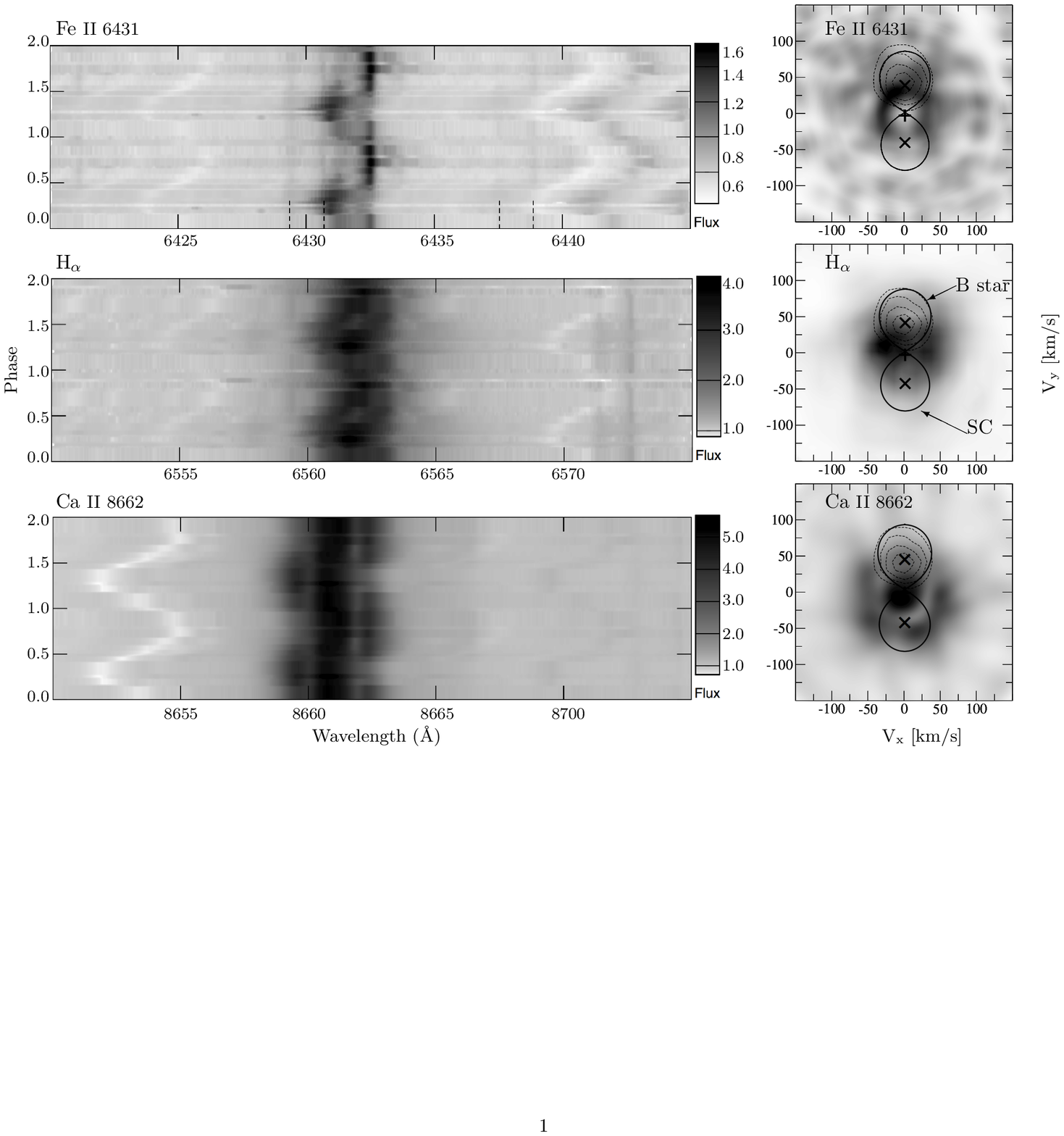}}}
\end{picture}}
\caption{
Right panels: Doppler maps of AS 386 reconstructed for the Fe {\sc ii} 6431 \AA\ (upper),
H$\alpha$ (middle), and Ca {\sc ii} 8662 \AA\ (bottom) emission lines.
Two crosses mark the positions of  the centers of mass of the components.
The plus corresponds to the position of the center of mass of the binary system.
Roche lobe boundaries (the thick lines) of B-type star (B star) and the secondary component (SC), the black hole candidate, are marked by the arrows in H$\alpha$ map, respectively.
The dashed-line contours show the Doppler maps of the absorption Si {\sc ii} 6347 \AA\ line.
Left panels: Trailed spectra in a region around the same lines folded with the orbital period. Intensity (relative flux with respect
to the local continuum) scale is shown on the right side, and the orbital phase is shown on the left side.
Short dashed lines in the top-left panel mark pairs of weak unidentified not moving lines (see Sect.\,\ref{tomography}). Intensities and wavelengths are shown in the same units as in Fig.\,\ref{f3}.
}
\label{f13}
\end{figure*}

If we assume that the secondary component is much hotter, then it should be much smaller in size than the B-type component and introduce a UV excess to the system SED. With a reasonable surface gravity of $\log$ g = 5.0 and a mass of $\sim$7.5 M$_{\odot}$, one can derive a radius of $\sim$1.4~R$_{\odot}$ for such a hot component.
Assuming its bolometric luminosity is equal to that of the B-type component, the hot star should have a T$_{\rm eff} \sim$36000 K and contribute $\sim$10 \% of the flux
to the $U$--band. Nevertheless, we do not see any excess of the $U$--band flux (see Sect.\,\ref{sed} and Fig.\,\ref{f11}). 

The 1988 $U-B$ color-index from
\citet{1989AJ.....98.1768C} is $\sim$0.4 mag more negative compared to our data. It might be a result of a large aperture size (25$\arcsec$) used in these observations.
If real, such a strong excess would require a higher luminosity of the hotter secondary that
would cause a stronger contribution to the system optical flux. Such a component would also cause periodic appearance of a hot spot on the surface of the B-type
component and thus introduce periodic variations of the observed line profiles which has not been detected either.

Hot helium subdwarfs (T$_{\rm eff} \ge 30000$ K) have been detected in the UV spectra of some Be binaries that underwent mass transfer, are typically much less massive and much fainter beyond the UV spectral range than their B-type components \citep[e.g.,][]{1995ApJ...448..878T}. Generally helium subdwarfs are stars with masses below
1 M$_{\odot}$ which are formed through various types of interaction in close binary systems \citep[e.g.,][]{2016MNRAS.463.2756H}.

The assumption that the secondary component is a main-sequence early B-type star with a current mass of 7 M$_\odot$ and a luminosity $\sim$3 times smaller than the
B-type primary (see Fig.\,\ref{f12}) is inconsistent with the observational data either.
Even if it is fast-rotating, which would make its spectral lines shallow and hard to detect on the background of the brighter primary, it would also introduce a $\sim$0.2 mag
excess in the $U-B$ color and leave the unusual metal abundance of the primary unexplained.
Therefore, we conclude that our data leave us the only opportunity to assume that the invisible secondary component is a black hole with a mass of $\gtrsim 7$~M$_{\odot}$.

\subsection{Dust temperature variations}\label{dust_temp}

The B-type component has such a T$_{\rm eff}$ that dust sublimation is only possible well beyond the orbits of the stellar components. The dust should be located at a
distance of $\sim$100 R$_{\rm hot}$, where R$_{\rm hot}$ is the radius of the B-type component \citep[see, e.g.,][for examples of the dust sublimation distance near hot stars]{1999ApJ...520L.115M,2006ApJ...636..348V}.

Since the B-type component moves around the center of mass of the system, it periodically changes its distance from the part of the inner rim of the dusty disk that faces
the observer. In particular, the star is the closest to that part at the orbital phase 0.0 (superior conjunction) and the farthest at the phase 0.5 (inferior conjunction).

The maximum height of the inner rim (H$_{\rm rim}$, measured from the disk midplane) is 20\% of its radius (R$_{\rm rim}$) in order to avoid shadowing large
portions of the disk \citep{2001ApJ...560..957D}.
We see no evidence for any additional reddening in the object's SED other than interstellar (see Sect.\,\ref{absorptions}).
Therefore, the system orbital plane has to be tilted with respect to the line of sight by a minimum angle of $\arctan$(0.5\,H$_{\rm rim}$/R$_{\rm rim}) \sim 11\degr$
(inclination 79$\degr$) to avoid obscuration of the star by the disk (ignoring the disk flaring).
The orbital solution implies a total mass of the system of at least 14 M$_\odot$ (see Fig.\,\ref{f6}) that
leads to a components separation of 1.22 AU. In turn, the orbital radius of the B-type component is half this separation or 8 R$_{\rm hot}$ (see Sect.\,\ref{absorptions}).

According to \citet{2006ApJ...636..348V}, the inner rim temperature (T$_{\rm rim}$) varies with the distance from the illuminating star (D$_{\star}$) as follows:
T$_{\rm rim} \propto$ D$_{\star}^{-0.5}$. For the adopted orbital radius and R$_{\rm rim}$ = 100\,R$_{\rm hot}$, T$_{\rm rim}$ changes by 8\% between phases 0.0 and 0.5.
Since dust emits as a blackbody, the amount of radiation it emits at different temperatures will vary with the Planck function \citep[equation 3 of][]{2006ApJ...636..348V}.
Assuming that T$_{\rm rim}$ is equal to a dust sublimation temperature of 1500 K, one can calculate that the inner rim flux at $\lambda = 2\,\mu$m ($K$--band) will
change by $\sim$40 \% (0.4 mag) due to the B-type component maximum illumination difference. This result is very close to the amplitude of the observed $K$--band
brightness variations (0.3 mag, see Table\,\ref{t2}) and thus qualitatively supports our hypothesis about their origin. Additionally, it can explain a higher intensity of the CO emission bands near the phase 0.0 (see Fig.\,\ref{f2}), when the dusty continuum is weaker.

Our assumption for the H$_{\rm rim}$ value is supported by the ratio of the observed fluxes at 2 and 1 $\mu$m, which is equal to 0.25$\pm$0.03 and was calculated using
the SED (Fig.\,\ref{f11}) accounting for the $K$--band brightness variations.
This ratio was suggested to be a measure of the strength of the near-IR excess due to the radiation of the disk inner rim and depends on its height,
because the observed flux at $\lambda = 1 \mu$m is dominated by the stellar radiation, while that at $\lambda = 2 \mu$m is dominated by the dust radiation.
A theoretical ratio of 0.23 for non-gray dusty particles \citep[from equation A10 of][]{2006ApJ...636..348V} for the adopted T$_{\rm eff}$ of the B-type component, dust sublimation temperature, and H$_{\rm rim}$/R$_{\rm rim}$ = 0.2 coincides with our ratio within the uncertainties.
A more detailed modeling of the circumstellar dust in AS\,386 requires data on its composition and is beyond the scope of this paper.

\subsection{Doppler tomography}\label{tomography}

We used Doppler tomography \citep{1988MNRAS.235..269M} to study formation regions of the H$\alpha$, Fe {\sc ii} 6431 \AA, and Ca {\sc ii} 8662 \AA\
emission lines and the Si {\sc ii} 6347 \AA\ absorption line.
The Fe {\sc ii} and Ca {\sc ii} lines were chosen, because they are among the strongest metallic lines in the optical spectrum.
Also, they are not contaminated by other lines and show different profiles. The Fe {\sc ii} line is double-peaked with both peaks moving with the B-type component, while
the Ca {\sc ii} line shows a stationary double-peaked structure with a moving unresolved central emission peak (see Fig.~\ref{f7}).

Figure\,\ref{f13} shows phased time series spectra around the Fe {\sc ii} 6431 \AA\ (top), H$_\alpha$ (middle), and Ca {\sc ii} 8662~\AA\ (bottom) lines (left panels) and
corresponding Doppler maps\footnote{The Doppler maps were built by combining the phase-resolved spectra using the maximum entropy method as implemented by
\citet{1998astro.ph..6141S} http://www.mpa-garching.mpg.de/$\sim$henk/pub/dopmap}
(right panels).  The inclination angle $i=70\degr$ is chosen based on its limiting value (see Sect.\,\ref{dust_temp}) and the observed RV amplitude of the B-type
component which implies that it cannot be very low.
The orbital period P$_{\mathrm {orb}}=131.3$\,days as well as the masses of B-type component  M$_{1}=7.0$\,M$_\odot$ and the black hole
M$_{2}=7.4$\,M$_\odot$  (the mass ratio $q\eqsim0.95$) are used to overlay the positions and the Roche lobes of the stellar components.
The orbital phase $\phi_{orb}$ = 0.0 corresponds to a superior conjunction of the B-type component.
Changing the inclination angle does not alter the tomography results and only affects the Roche lobe sizes and the mass of the black hole.
The dashed-line contours show the tomography constructed using the Si~{\sc ii} 6347 \AA\ absorption line. This line is originated in the atmosphere of the B-type
component.

The regions of formation of the emission lines have a more complicated structure. We clearly see a ring around the center of mass of the system located at a velocity of $\sim$$30~$km\,s$^{-1}$ in both the Fe {\sc ii} 6431 \AA\ and  Ca {\sc ii} 8662 \AA\ lines.  The intensity of the ring in Fe {\sc ii} 6431 \AA\  seems to be non-uniform with a maximum  at ($V_x \approx -10~\mathrm{ km~s}^{-1}$, $V_y \approx 30~\mathrm{ km~s}^{-1}$) located close to the velocity of the B-type component ($V_x = 0~\mathrm{ km~s}^{-1}$, $V_y = 51~\mathrm{ km~s}^{-1}$) and a minimum at the opposite side.
We suggest that the ring corresponds to the inner boundary of the circumbinary disk which is most likely perturbed by the wind from the B-type component.

Three components can be distinguished in the Ca {\sc ii} 8662~\AA\ map: (1)  a bright compact region around the zero velocity close to the inner  Lagrangian point, L$_1$, or/and the system center of mass; (2) a nonuniform ring at $V \sim 50~\mathrm{ km~s}^{-1}$;  (3) a visible deficit of emission at the position of the Fe {\sc ii} 6431 \AA\ ring which
is probably caused by a superposition of the flux from the compact zero velocity region and circumbinary disk in the
Ca {\sc ii} 8662~\AA\ line, which shows a slightly higher velocity compared with that of the Fe {\sc ii} line.

The circumbinary disk structure is not well detected in the H$\alpha$ line Doppler map. It shows two distinct components: an extended bright spot at ($V_x \approx -20~\mathrm{ km~s}^{-1}$, $V_y \approx 10~\mathrm{ km~s}^{-1}$) and a fainter structure at the opposite side with ($V_x \approx 30~\mathrm{ km~s}^{-1}$, $V_y \approx 0~\mathrm{ km~s}^{-1}$). The first component is more likely a shock region between the stellar wind from the B-type component and the circumbinary disk,
and the second one is a visible residue of the circumbinary disk in this map.

We note the presence of paired non-moving weak lines, which are hard to identify by inspecting the spectrum. Some of them are marked by the dashed lines
in the Fe~II 6431 \AA\ region trailed spectra in Fig.\,\ref{f13} (top left panel). Such lines are also present in the H$\alpha$ region (Fig.\,\ref{f13}, middle left panel).
These double-peaked lines probably form in the circumbinary disk and can be identified with low-excitation species (e.g., Ca {\sc i} 6572.78 \AA).

\section{Discussion}\label{discussion}

Analysis of our data shows that AS\,386 is a single-lined binary system on a circular orbit with a period of 131.3~days.
The B-type component parameters along with the orbital solution imply that the stars currently have comparable, intermediate masses. They are located at $\sim$1.2 AU from one another and should have very different luminosities. The B-type component is responsible for the optical radiation from the system, while the circumstellar dust dominates
the IR region. Overall, AS\,386 fits the definition of a FS\,CMa type object \citep{2007ApJ...667..497M}. Although the B-type component is formally classified as a supergiant,
its luminosity is well below the upper boundary for the group luminosity ($\log$ L/L$_{\odot} = 4.5$). There is growing evidence that Galactic B[e] supergiants are binary
systems, which are just more massive than FS\,CMa type objects \citep[see, e.g.,][]{2012A&A...545L..10W}. Therefore, both groups can be products of the same
evolutionary channel, and dusty environments of FS\,CMa objects may survive for a longer time due to their weaker stellar winds. Close investigation of many
still poorly studied members and candidates to both groups will shed more light on this subject.

The position of the B-type component in the Hertzsprung-Russell diagram is typical for a mass derived from evolutionary tracks for both single stars and interacting binaries, but the secondary component has too low a luminosity for the mass expected from the system mass function.
As was shown by \citet{2001A&A...369..939W}, components of interacting binaries may stay over- or under-luminous for their current mass by up to a factor of 2 but only for a short period of time during the active mass transfer. We see no evidence for a Roche lobe overflow in the AS\,386 binary system. Both components should be confined well inside their Roche lobes. The secondary component's mass is no lower than $\sim$7 M$_{\odot}$ even for an edge-on orientation of the orbital plane.
Therefore we have to assume that it is a black hole. No X-ray or $\gamma$-ray source has been detected near the position of AS\,386 in the sky, but the large components' separation and the weak emission-line spectrum, which suggests a relatively small amount of circumstellar gas within the components' Roche lobes, lead us to a conclusion that the black hole cannot presently reveal itself through accretion of the circumstellar material.

The presence of numerous absorption lines of such elements as silicon, neon, and aluminum in the optical spectrum of the B-type component is unusual
(see, e.g., bottom panels of Fig.\,\ref{f3} and \ref{f4}).
Spectra of normal stars of any luminosity type and similar T$_{\rm eff}$ show very weak lines between H$\alpha$ and the first telluric band near $\lambda \sim$ 6800 \AA\
except for those of He {\sc i} and DIBs. This can be explained by the enrichment of the B-type component's atmosphere with the material from the more evolved
secondary component and perhaps with those of its final explosion. The absence of a nebula around AS\,386 suggests that the explosion occurred long ago, and the
B-type component significantly evolved after accepting the material from the secondary component. As seen in Fig.\,\ref{f12}, it is now the most evolved object
among the FS\,CMa group members which are supposed to follow the same evolutionary scenario \citep[mass transfer in close binaries, e.g.,][]{2017ASPC..508..285M}.

The presence of a massive compact component (a neutron star or a black hole) in a binary system implies that it was initially much more massive than the second component, but its explosion as a supernova has not disrupted the system. It was suggested for a recently discovered binary with a Be star and a black hole, MWC\,656 \citep{2015MNRAS.452.2773G}, that the initial mass of the more massive component was $\sim$41 M$_{\odot}$. Another example of a similar system is CI\,Cam, an object with the B[e] phenomenon, whose multiwavelength outburst in 1998 revealed the presence of a compact component near an early B--type star \citep[e.g.,][]{2000A&A...356...50C}. It is still unclear what kind of compact component exists in this system, but it was probably more massive than the presently $\sim$10 M$_{\odot}$ visible star.

In both mentioned cases, the compact component reveals itself by the presence of a weak He {\sc ii} 4686 \AA\ emission line that traces the orbit of this component by
regularly changing its radial velocity. Both these systems have shorter orbital periods (19.4 days in CI\,Cam and 60 days in MWC\,656) and stronger emission-line
spectra than that of AS\,386.

\section{Conclusions}\label{conclusions}

Our spectroscopic and photometric study of the emission-line type object AS\,386 revealed another binary system among the FS\,CMa group members.
We found that it is a single-lined binary on a circular orbit with a period of 131.27 days and a RV semi-amplitude of 51.8 km\,s$^{-1}$. The component responsible
for the radiation in the visible region is a late B-type star with T$_{\rm eff} = 11000\pm500$ K,  $\log$ L/L$_{\odot} = 3.7\pm0.3$, and R = $20\pm5$ R$_{\odot}$.
Its current mass estimated from the evolutionary tracks is $7\pm1$ M$_{\odot}$ implying a surface gravity of $\log$ g $= 2.7\pm0.3$.
The system is located at a distance of $D = 2.4\pm0.3$ kpc from the Sun and affected by a reddening of E($B-V$) = 0.85$\pm$0.03 mag, which is found to
be totally interstellar.

No traces of the secondary component have been found in our high-resolution optical ($\lambda\lambda$ 3600--10500 \AA) and low-resolution near-IR
($\lambda\lambda$ 1--2.5 $\mu$m) spectra. Also, no convincing evidence for eclipses has been found in the brightness variations of the system.
The mass function derived from the orbital solution, $f(m) = 1.9\pm0.3$ M$_{\odot}$, combined with the B-type component
mass leads to a mass of $\ge 7$ M$_{\odot}$ for the secondary component irrespective of the orbital inclination angle. We suggest that the system properties can
only be explained by assuming that the massive secondary is a black hole. This conclusion is indirectly supported by the presence of an unusual number of
absorption lines, which unambiguously belong to the B-type component's atmosphere. The material responsible for these lines might have come from the initially
more massive secondary component during the mass transfer phase and its final explosion. The amount of the circumstellar material in the AS\,386 system is not
large enough to build a gaseous disk around the secondary component and thus reveal its presence.

The emission lines exhibit a complex structure due to formation in several regions. One of them is a circumbinary disk, where both the forbidden (e.g., [Ca {\sc ii}])
and permitted (e.g., Fe {\sc ii}) lines are stationary. Their double-peaked profiles regularly vary due to the orbital motion of the B-type component, which provides
a variable excitation to the disk material. Another region is most likely located within the binary, near its center of mass, and is manifested by the extremely strong
(even stronger than H$\alpha$) emission lines of the Ca {\sc i} triplet (8498, 8542, 8662 \AA).
They are the strongest among nearly 30 FS\,CMa objects observed in this spectral region and might be indicative of the history of the mass transfer or the
evolutionary status of the system.

The orbital motion of the B-type component also affects the brightness of the inner rim of the circumbinary dusty disk that varies by $\sim$30 \% in the $K$--band.
We qualitatively showed that it is a pure effect of the variable rim temperature due to the illuminating flux variations. The wavelength dependence of the IR excess
in AS\,386 is very similar to those of a number of FS\,CMa objects \citep[e.g., MWC\,728,][]{2015ApJ...809..129M}. However, since we do not have information
about the dust composition (no mid-IR spectra or interferometry of AS\,386 has been taken so far), modeling of its dusty environment is premature. At the same time,
it is worth looking for similar near-IR brightness variations in other dusty binaries to better constrain the properties of the dust.
The observed variations of the optical brightness can be due to an inhomogeneous distribution of the circumstellar matter which is hinted in the results of the Doppler
tomography (see right panels of Fig.\,\ref{f13}).

AS\,386 is the third FS\,CMa type binary with determined orbit after CI\,Cam \citep{2006ASPC..355..305B} and MWC\,728 \citep{2015ApJ...809..129M}.
It may also be the first one with a black hole secondary component, if this is not proven for CI\,Cam. In any case, both these objects seem to have a compact
component in their systems suggesting that circumstellar dust can exist around hot stars for a very long time. This reinforces earlier suggestions that FS\,CMa
type objects can be a noticeable source of the Galactic dust. \\

\acknowledgements
A.~M. and S.~Zh. acknowledge support from DGAPA/PAPIIT Project IN\,100617.
This research has made use of the SIMBAD database, operated at CDS, Strasbourg, France.
This paper is partly based on observations obtained with the Canada-France-Hawaii Telescope (CFHT) which is operated by the
National Research Council of Canada, the Institut National des Sciences de l$^{\prime}$Univers of the Centre National de la
Recherche Scientifique de France, and the University of Hawaii as well as on observations obtained at the 2.7\,m Harlan J. Smith
telescope of the McDonald Observatory (Texas, USA), the Apache Point Observatory 3.5\,m telescope, which is owned and operated
by the Astrophysical Research Consortium, and the 2.1\,m of the Observatorio Astron\'omico Nacional San Pedro Martir (Baja California, M\'{e}xico).
The observations at the Canada-France-Hawaii Telescope were performed with care and respect from the summit of Maunakea which is a significant
cultural and historic site. The observations at the Tien-Shan Astronomical Observatory were carried out thanks to the support from the program BR05236322
of the Ministry of Education and Science of the Republic of Kazakhstan.
We are grateful to the 1st and 2nd year NMSU astronomy graduate students for acquiring the ARCES spectrum on MJD 58070.\\
\bibliography{AS386_ApJ_arxiv}
\newpage

\setcounter{table}{3}
\begin{table*}                                          
\begin{center} 
\caption{List of lines identified in the spectrum of AS 386}
\begin{tabular}{lrcrrl|lrcrrl}
\hline\noalign{\smallskip}
Line         & Mult. &$\lambda_{\rm lab}$&EW & I/Ic & Comment &                    Line         & Mult. &$\lambda_{\rm lab}$&EW & I/Ic & Comment \\
\noalign{\smallskip}\hline\noalign{\smallskip}\noalign{\smallskip}\hline\noalign{\smallskip}
He {\sc i}	     &  22	   &3819.61	&0.70	&0.55 	&                                        &  Fe {\sc ii}	&27	   &4416.82	&0.11	&0.74 	&     \\         
He {\sc i}	     &  62	   &3833.57	&0.33	&0.71 	&                                        &  Mg {\sc ii}	&9	   &4428.00	&0.12	&0.78 	&     \\         
H  {\sc i}       &  2	   &3835.39	&0.30	&0.66	&    H9                                  &  S  {\sc ii}	&32	   &4431.02	&0.11	&0.76 	&     \\      
Mg {\sc i}       &  3 	   &3838.29      &0.09 	&0.86 	&                                    &  S  {\sc ii}	&43	   &4432.41	&0.08	&0.82 	&     \\             
S  {\sc ii}      &  22	   &3845.21	&0.08	&0.80 	&                                        &  Mg {\sc ii}	&9	   &4433.99	&0.10	&0.77 	&     \\         
Mg {\sc ii}	     &  5	   &3848.24	&0.08	&0.80 	&                                        &  Mg {\sc ii}	&19	   &4436.48	&0.36	&0.77 	&+f   \\         
Si {\sc ii}	     &  1	   &3853.66	&0.28	&0.61 	&                                        &  He {\sc i}	&50	   &4437.55	&0.36	&0.72 	&+p   \\         
Si {\sc ii}	     &  1	   &3856.02	&0.36	&0.49 	&                                        &  Fe {\sc ii}	&187   &4446.25	&0.03	&0.91 	&     \\       
Si {\sc ii}	     &  1	   &3862.59	&0.35	&0.45 	&                                        &  N  {\sc ii}	&15	   &4447.03	&0.06	&0.82 	&     \\         
He {\sc i}	     &20	   &3867.63	&0.24	&0.61 	&                                        &  Fe {\sc ii}	&	   &4455.26	&0.04	&0.87 	&     \\         
He {\sc i}       &60	   &3871.82	&0.55	&0.68 	&                                        &  S  {\sc ii}	&43	   &4456.43	&0.03	&0.95 	&     \\         
H  {\sc i}       & 2 	   &3889.05	&0.57	&0.63 	&  H8                                    &  S  {\sc ii}	&43	   &4463.58	&0.07	&0.80 	&    \\      
S  {\sc ii}	     &50	   &3892.32	&0.09	&0.86 	&                                        &  S  {\sc ii}	&	   &4464.43	&0.05	&0.80 	&    \\         
Ti {\sc ii}	     &34	   &3900.55	&0.13	&0.81 	&                                        &  He {\sc i}	&14	   &4471.48	&0.40	&0.50	 &    \\         
Ti {\sc ii}	     &34	   &3913.46	&0.10	&0.85	&                                        &  Mg {\sc ii}	&4	   &4481.23	&0.49	&0.47 	&    \\         
C {\sc ii}       &4	          &3920.68	&0.11	&0.85 	&                                        &  S  {\sc ii}	&43	   &4483.42	&0.10	&0.78 	&    \\             
S {\sc ii}       &55	   &3923.48	&0.14	&0.84 	&                                        &  Fe {\sc ii}	&38	   &4491.40	&0.13	&0.73 	&    \\         
He {\sc i}	     &58	   &3926.53	&0.32	&0.72	&                                        &  Fe {\sc ii}	&38	   &4508.28	&0.15	&0.69	 &    \\         
Ca {\sc ii}      &1 	   &3933.66	&1.20 	&0.34 	&                                        &  S  {\sc ii}	&48	   &4509.00	&0.05	&0.87 	&    \\         
He {\sc i}	     &5	       &3964.73	&0.38	&0.64 	&                                        &  Fe {\sc ii}	&37	   &4515.34	&0.12	&0.75 	&    \\        
H  {\sc i}       &1	            &3970.07	&1.27	&0.41 	&  H$\epsilon$+Ca {\sc ii}(1)            &  Fe {\sc ii}	&38	   &4522.63	&0.16	&0.69	 &    \\        
S {\sc ii}       &45	   &3990.94	&0.09	&0.88 	&                                        &  Fe {\sc ii}	&37	   &4534.17	&0.11	&0.77 	&    \\        
S {\sc ii}       &29	   &3993.53	&0.12	&0.83 	&                                        &  Fe {\sc ii}	&38	   &4541.52	&0.08	&0.81 	&    \\        
K {\sc ii}       &1	          &3995.10	&0.30	&0.66 	&                                        &  N  {\sc ii}	&58	   &4552.54	&0.21	&0.70 	&  + Si {\sc iii} (2)   \\  
S {\sc ii}       &59	   &3998.79	&0.09	&0.88 	&                                        &  Cr {\sc ii}	&44	   &4555.02	&0.04	&0.88 	&     		\\ 		      
He {\sc i}       &55	   &4009.27	&0.44	&0.60 	&                                        &  Fe {\sc ii}	&37	   &4555.89	&0.13	&0.71 	&           \\               
He {\sc i}       &18	   &4026.36	&0.62	&0.50 	&                                        &  Cr {\sc ii}	&44	   &4558.66	&0.14	&0.69 	&           \\                  
S {\sc ii}     	 &59	   &4032.81	&0.11	&0.86 	&                                        &  Ar {\sc ii}	&	   &4563.78	&0.07	&0.81 	&           \\               
H$\delta$        &1	       &4101.74	&0.45	&0.64	&                                        &  Si {\sc iii}&2 	   &4567.87	&0.11	&0.82 	&           \\               
He {\sc i}       &16	   &4120.81	&0.32	&0.66 	&                                        &  Si {\sc iii}&2     &4574.78	&0.07	&0.82 	&           \\               
Si {\sc ii}      &3	       &4128.05	&0.24	&0.73 	&                                        &  Fe {\sc ii}	&38	   &4576.33	&0.09	&0.76 	&           \\               
Si {\sc ii}      &3	       &4130.88	&0.30	&0.66 	&                                        &  Fe {\sc ii}	&38	   &4583.83	&0.18	&0.61 	&           \\               
S {\sc ii}       &44	   &4142.29	&0.13	&0.78 	&                                        &  Cr {\sc ii}	&44	   &4588.22	&0.14	&0.68 	&           \\               
He {\sc i}       &53	   &4143.76	&0.54	&0.55 	&                                        &   Cr {\sc ii}	&44	   &4589.89	&0.05	&0.84 	&       \\                   
S {\sc ii}       &44	   &4145.10	&0.20	&0.79	&                                        &   N  {\sc ii}	&20	   &4803.27	&0.05	&0.88 	&       \\                   
S {\sc ii}       &65	   &4146.94	&0.07	&0.90 	&                                        &   N  {\sc ii}	&20	   &4810.29	&0.07	&0.90 	&        \\                  
S {\sc ii}       &44	   &4162.70	&0.15	&0.79 	&                                        &   Cr {\sc ii}	&30	   &4812.35	&0.02	&0.92 	&        \\                  
S {\sc ii}       &44	   &4168.41	&0.08	&0.85 	&                                        &   S  {\sc ii}	&9	   &4815.52	&0.27	&0.60 	&        \\                  
He {\sc i}     	 &52	   &4168.97	&0.13	&0.80 	&                                        &   S  {\sc ii}	&46    &4819.60	&0.08	&0.84 	&        \\                  
S {\sc ii}     	 &67	   &4230.98	&0.08	&0.89 	&                                        &   S  {\sc ii}	&52	   &4824.07	&0.09	&0.80 	&        \\                  
Fe {\sc ii}      &27	   &4233.17	&0.23	&0.61 	&                                        &   Cr {\sc ii}	&30	   &4824.13	&0.10	&0.79 	&        \\                  
S {\sc ii}       &66	   &4259.18	&0.08	&0.83 	&                                        &   Fe {\sc ii}	&30	   &4833.21	&$-$0.10&1.09	& dp     \\                  
C {\sc ii}       &6	       &4267.02	&0.27	&0.74 	&+f                                      &   Cr {\sc ii}	&30	   &4836.22	&0.03	&0.90	&        \\             
S {\sc ii}       &49	   &4267.80	&0.27	&0.75 	&+p                                     & Cr {\sc ii}	&30	   &4848.24	&0.05	&0.83 	&            \\         
S {\sc ii}       &49	   &4294.43	&0.21	&0.75 	&                                      &Mg {\sc ii}	&25	   &4851.10	&0.14	&0.80	&                \\     
Fe {\sc ii}      &28	   &4296.57	&0.12	&0.75 	&                                     &Cr {\sc ii}	&30	   &4856.19	&0.03	&0.92 	&                \\     
Mg {\sc ii}	     &27	   &4331.93	&0.09	&0.83 	&                                     &H$\beta$	&1	   &4861.33	&$-$0.18&1.17 	& dp                 \\
H$\gamma$   &1	    &4340.47     &0.33	&0.65 	&                                      &Cr {\sc ii}	&30	   &4864.32	&0.05	&0.86 	&                \\    
Mn {\sc ii}      &6	    &4343.99	&0.03	&0.88 	&                                     &Cr {\sc ii}	&30	   &4876.41	&0.08	&0.75 	&                \\    
Ar {\sc ii}	     &7	    &4348.11	&0.08	&0.88 	&                                      &Cr {\sc ii}	&30	   &4884.57	&0.03	&0.90 	&            \\        
Fe {\sc ii}	     &27	   &4351.76	&0.12	&0.65 	&                                     &S  {\sc ii}	&15	   &4885.63	&0.12	&0.75 	&                \\    
Fe {\sc ii}	     &27	   &4357.57	&0.04	&0.87	 &                                     &Fe {\sc ii}	&36	   &4893.78	&$-$0.14&1.09	        & dp     \\    
Fe {\sc ii}	     &27	   &4361.25	&0.03	&0.88 	&                                     &Fe {\sc ii}	&	   &4908.15	&0.08	&0.83 	&                \\    
Ar {\sc ii}	     &39	   &4362.07	&0.06	&0.81	 &                                      & Fe {\sc ii}	&	   &4913.30	&0.10	&0.80	&            \\
Fe {\sc ii}	     &             &4369.40	&0.03	&0.88 	&                                        & S  {\sc ii}	&15	   &4917.15	&0.15	&0.71 	& \\
Ar {\sc ii}	     &1	       &4371.36	&0.05	&0.89 	&                                      &He {\sc i}	&48	   &4921.93	&0.34	&0.59 	&                   \\
Mg {\sc ii}	     &10	   &4384.64	&0.04	&0.81 	&                                      & S  {\sc ii}	&7	   &4925.32	&0.19	&0.69 	&                   \\
Fe {\sc ii}	     &27	   &4385.38	&0.11	&0.72 	&                                       & S  {\sc ii}	&7	   &4942.47	&0.10	&0.83 	&                    \\
He {\sc i}	    &51	   &4387.93	&0.58	&0.46 	&                                        &Fe {\sc ii}	&	   &4951.59	&0.11	&0.79 	&                    \\
Mg {\sc ii}     &10	   &4390.59	&0.10	&0.70 	&                                      &Cr {\sc ii}	&	   &4952.78	&0.04	&0.91 	&                    \\
Fe {\sc iii}    &4   	   &4395.78	&0.10	&0.78 	&                                      &Fe {\sc ii}	&168   &4953.98	&0.02	&0.93 	&                    \\
\noalign{\smallskip}\hline\noalign{\smallskip}
\end{tabular}
\end{center}
\end{table*}

\setcounter{table}{3}
\begin{table*}
\begin{center}
{\bf Table 4. continued: }{ List of lines identified in the spectrum of AS 386}
\begin{tabular}{lrcrrl|lrcrrl}
\hline\noalign{\smallskip}
Line         & Mult. &$\lambda_{\rm lab}$&EW & I/Ic & Comment &      Line         & Mult. &$\lambda_{\rm lab}$&EW & I/Ic & Comment \\                
\noalign{\smallskip}\hline\noalign{\smallskip}\noalign{\smallskip}\hline\noalign{\smallskip}
Fe {\sc ii}	&318   &4957.58	&$-$0.01&1.06	& dp                 & O  {\sc i}	&12	   &5329.50	&0.31	&	         & +f   \\
Fe {\sc ii}	&	   &4984.50	&0.11	&0.77 	&                    & O  {\sc i}	&12	   &5330.74	&0.31	&	         & +p   \\
Fe {\sc ii}	&	   &4990.50	&0.07	&0.82	&                    & Cr {\sc ii}	&43	   &5334.87	&0.03	&0.89 	       &    \\
S {\sc ii}	&7	   &4991.94	&0.16	&0.72 	&                    & Fe {\sc ii}	&48	   &5337.73	&$-$0.14	&1.13	& dp \\
Fe {\sc ii}	&	   &4993.35	&$-$0.18&1.16 	& dp                 & Fe {\sc ii}	&	   &5339.59	&0.19	&0.76  	      &    \\
N {\sc ii}	&24    &4994.36	&0.03	&0.92 	&                    & S  {\sc ii}	&38	   &5345.72	&0.20	&0.75	        &    \\
Fe {\sc ii}	&	   &5004.20	&0.72	&0.72 	&                    & Fe {\sc ii}	&48	   &5362.86	&$-$0.11	&1.16	 & dp \\
N {\sc ii}	&19    &5005.15	&0.08	&0.83 	&                    & Fe {\sc ii}	&	   &5370.30	&0.07	&0.87  	      &    \\
S {\sc ii}	&7	   &5009.54	&0.26	&0.68 	&                    & Fe {\sc ii}	&	   &5387.07	&0.16	&0.75  	      &    \\
S {\sc ii}	&15	   &5014.03	&0.75	&0.66 	&+f                  & Fe {\sc ii}	&	   &5395.86	&0.07	&0.86 	       &    \\
He {\sc i}	&4	   &5015.68	&0.75	&0.75 	&+p                  & Fe {\sc ii}	&184   &5408.81	&$-$0.11  	&1.11	 & dp \\
Fe {\sc ii}	&42	   &5018.44	&$-$0.10&1.21 	&   				   Fe {\sc ii}	&48	   &5414.06	&$-$0.14 	&1.14	& dp \\
S {\sc ii}	&7	   &5032.41	&0.37	&0.57 	&                    & Fe {\sc ii}	&49	   &5425.25	&$-$0.15	&1.18	& dp  \\
Fe {\sc ii}	&	   &5035.71	&0.17	&0.70 	&                    & S  {\sc ii}	&6	   &5428.67	&0.22	&0.68	 &    \\
Si {\sc ii}	&5	   &5041.06	&0.17	&0.71 	&                    & Fe {\sc ii}	&	   &5429.99	&0.14	&0.78	 &    \\
N {\sc ii}	&4	   &5045.10	&0.18	&0.71 	&                    & S  {\sc ii}	&6	   &5432.82	&0.14	&0.78	 &    \\
Fe {\sc ii}	&	   &5047.64	&0.37	&0.65 	&                    & S  {\sc ii}	&6	   &5453.83	&0.41	&0.52	 &    \\
He {\sc i}	&47	   &5047.74	&0.36	&0.66 	&                    & Fe {\sc ii}	&	   &5457.72	&0.07	&0.85	 &    \\
Si {\sc ii}	&5	   &5056.02	&0.26	&0.67 	&                    & S  {\sc ii}	&6	   &5473.62	&0.30	&0.63	 &    \\
Fe {\sc ii}	&	   &5061.73	&0.13	&0.76 	&                    & Fe {\sc ii}	&	   &5482.31	&0.12	&0.74	 &    \\
Fe {\sc ii}	&	   &5067.89	&0.06	&0.87 	&                    & Fe {\sc ii}	&	   &5487.63	&0.20	&0.73	 &    \\
Fe {\sc ii}	&	   &5070.90	&0.13	&0.76 	&                    & DIB	&	   &5494.10	&0.13	&0.77 	&    \\
Fe {\sc iii}&5	   &5073.90	&0.10	&0.84	&                    & Fe {\sc ii}	&	   &5506.20	&0.31 	   &0.73 	&    \\
Fe {\sc ii}	&	   &5075.77	&0.12	&0.77	&                    & S  {\sc ii}	&6	   &5509.72	&0.21	&0.68	 &    \\
Fe {\sc ii}	&	   &5082.23	&0.10	&0.83 	&                    & $[$Fe {\sc ii}$]$          	&56 	&5525.14	&$-$0.15 	&1.12 	& dp  \\
Fe {\sc ii}	&	   &5089.22	&0.11	&0.81	&                    & Fe {\sc ii}	&	   &5529.06	&0.05	&0.90	 &    \\
Fe {\sc ii}	&	   &5093.56	&0.14	&0.77 	&                    & Fe {\sc ii}	&224       	&5529.93	&$-$0.05	&1.08	& dp  \\
Fe {\sc ii}	&	   &5097.27	&0.11	&0.76 	&                    & Fe {\sc ii}	&	   &5532.09	&0.04	&0.89	 &    \\
Fe {\sc ii}	&35   	&5100.74&0.15	&0.69	&                    & Fe {\sc ii}	&	   &5534.84	&$-$0.36	&1.23	 & dp   \\
S  {\sc ii}	&7	   &5103.30	&0.17	&0.70 	&                    & Fe {\sc ii}	&	   &5548.21	&0.16	&0.88 	& +f   \\
Fe {\sc ii}	&	   &5117.03	&0.10	&0.83	&                    & Fe {\sc ii}	&	   &5549.00	&0.16	&0.86	 & +p   \\
$[$Fe {\sc ii}$]$&35&5120.34&$-$0.09&1.13	&                    & O {\sc i}	&24	   &5555.00	&0.04	&0.92	 &    \\
Fe {\sc ii}	&	   &5123.19	&0.03	&0.95 	&                    & S {\sc ii}	&6	   &5556.01	&0.10	&0.83	 &    \\
Fe {\sc iii}&5	   &5127.35	&0.18	&0.81 	&                    & S {\sc ii}	&11	   &5578.89	&0.12	&0.86	 &    \\
$[$Fe {\sc ii}$]$&35&5132.67&$-$0.13&1.18	&                    & Fe {\sc ii}	&	   &5588.23	&0.10	 &0.82	 &    \\
S {\sc ii}	&1	   &5142.33	&0.11	&0.85 	&                    & Al {\sc ii}	&16	   &5593.30	&0.06	&0.89	  &    \\
Fe {\sc iii}&5	   &5156.12	&0.22	&0.79 	&                    & S {\sc ii}	&11	   &5606.15	&0.28	 &0.60	 &    \\
$[$Fe {\sc ii}$]$&18&5158.00&$-$0.18&1.18	&                    & S {\sc ii}	&11	   &5616.64	&0.13	 &0.78 	&    \\
Fe {\sc ii}	&	   &5177.30	&0.19	&0.78 	&                    & $[$Fe {\sc ii}$]$         	&57 	 &5627.49	&$-$0.12	&1.09	& dp  \\
Fe {\sc ii}	&49	   &5197.58	&$-$0.42&1.22	& dp                 & S  {\sc ii}	&4      	&5640.10	&0.52	 &0.58	 &    \\
Fe {\sc ii}	&	   &5199.12	&0.11	&0.82 	&                    & Fe {\sc ii}	&	   &5645.40	&0.51	&0.78	  & +f   \\
S {\sc ii}	&39	   &5201.16	&0.22	&0.55 	&                    & S {\sc ii}	&14	   &5647.03	&0.51	 &0.63	 & +p   \\
Si {\sc ii}	&	   &5202.51	&0.07	&0.87 	&                    & Fe {\sc ii}	&	   &5648.90	&0.10 	&0.85	  &    \\
Fe {\sc ii}	&	   &5203.64	&0.09	&0.81 	&                    & S {\sc ii}	&	   &5659.99	&0.26  	   &0.70 	&    \\
S {\sc ii}	&39	   &5212.61	&0.13	&0.70 	&                    & S {\sc ii}	&	   &5664.78	&0.21	 &0.72 	&    \\
Fe {\sc ii}	&	   &5227.49	&0.20	&0.72 	&                    & N  {\sc ii}	&3	   &5666.63	&0.20	 &0.74 	&    \\
Fe {\sc ii}	&49	   &5234.62	&$-$0.12&1.11	& dp                 & Si {\sc ii}	&	   &5669.56	&0.57	 &0.85 	& +f   \\
Cr {\sc ii}	&43	   &5237.32	&0.05	&0.86 	&                    & N  {\sc ii}	&3	   &5676.02	&0.57	 &0.75	 & +p   \\
Fe {\sc ii}	&	   &5247.95	&0.15	&0.76 	&                    & N  {\sc ii}	&3	   &5679.56	&0.28	 &0.66	 &    \\
Fe {\sc ii}	&	   &5251.23	&0.15	&0.75 	&                    & Na {\sc i}	&6	   &5682.63	&0.03	&0.95	  &    \\
Fe {\sc ii}	&49	   &5254.93	&$-$0.20&1.16 	& dp                 & N  {\sc ii}	&3	   &5686.21	&0.09	 &0.84	 &    \\
Fe {\sc ii}	&	   &5260.26	&0.16   &0.73	 &                   & Na {\sc i}	&6	   &5688.20	&0.12	&0.86	  &    \\
Mg {\sc ii}&17	   &5264.14	&0.04	 &0.89 	&                    & Fe {\sc ii}	&	   &5691.00	&0.06	 &0.89 	&    \\
Fe {\sc ii}	&	   &5291.67	&0.17	&0.74 	&                    & Al {\sc iii}    	&2	   &5696.60	&0.15	 &0.79 	&    \\
Cr {\sc ii}	&43	   &5308.44	&0.04	&0.88 	&                    & Si  {\sc ii}    	&	   &5701.38	&0.04	 &0.92 	&    \\
Cr {\sc ii}	&43	   &5310.70	&0.04	&0.90 	&                    & N  {\sc ii}	&3	   &5710.76	&0.13	&0.83	  &    \\
Cr {\sc ii}	&43	   &5313.58	&0.03	&0.91  	&                    & DIB     	&	   &5719.30	&0.10	 &0.92	&    \\
Fe {\sc ii}	&48    &5316.65	&$-$0.26&1.16	& dp                 &  Al {\sc iii}    	&2	   &5722.73	&0.08	 &0.86 	&    \\
S  {\sc ii}	&38	   &5320.73	&0.34	&0.65   &                    & Ne {\sc i}	&13	   &5764.42	&0.11	&0.85	  &    \\
Fe {\sc ii}	&49	   &5325.56	&$-$0.07&1.13	& dp                 & DIB	&	   &5780.37	&0.52	 &0.61	 &    \\
O  {\sc i}	&12	   &5328.98	&0.31	&	    & +f                 & Si {\sc ii}     	&8	   &5800.48	&0.12	 &0.86 	&\\
 \noalign{\smallskip}\hline\noalign{\smallskip}

\end{tabular}
\end{center}
\end{table*}

\setcounter{table}{3}
\begin{table*}
\begin{center}

\begin{center}
{\bf Table 4. continued: }{ List of lines identified in the spectrum of AS 386}
\end{center}
\begin{tabular}{lrcrrl|lrcrrl}
\hline\noalign{\smallskip}
Line         & Mult. &$\lambda_{\rm lab}$&EW & I/Ic & Comment &      Line         & Mult. &$\lambda_{\rm lab}$&EW & I/Ic & Comment \\  
\noalign{\smallskip}\hline\noalign{\smallskip}
 Ne {\sc i}	     &19   &5804.45	&0.08	&0.92 	 &                     & Si {\sc ii}                  	&	&6660.49  	&0.20 	&0.80 	& \\
Si {\sc ii}    	 &8	   &5806.75	&0.08	&0.85 	 & 					   & Si {\sc ii}                  	&  	&6665.03 	&0.08 	&0.90	& \\		
S {\sc ii}       &14   &5819.22	&0.21	&0.82 	 & +f                  & Si {\sc ii}                 	&  	&6671.88	 &0.32  	&0.68	& \\
Ne {\sc i}       &19   &5820.16	&0.21	&0.89  	& +p                   & Ne {\sc i}      	&6	   &6678.28     	&0.36 	&0.67	& + He {\sc i} 46 \\
Ne {\sc i}     	 &6	   &5852.49	&0.21	&0.72  	&                      & Fe {\sc ii}                  	& 	&6708.88 	&0.32 	&0.68 	& \\
Si {\sc ii}      &8	   &5867.50	&0.26	&0.82  	& +f                   & Ne {\sc i}                 	 &  	&6717.04  	&0.20	&0.76 	& \\
Si {\sc ii}      &8	   &5868.40	&0.26	&0.77 	 & +p                  & Ne {\sc i}               	&6	& 6929.47	&0.31         	&0.73	&\\
Ne {\sc i}       &1	   &5881.90	&0.22	&0.71  	&                      & Ne {\sc i}               	&6	&7024.05	&0.05       	&0.95  	& \\
Na {\sc i}       &11   &5889.95	&1.24 	&0.06	&                      & Ne {\sc i}                      	&1 	&7032.41 	&0.43 	&0.60 	& \\
Na {\sc i}       &11   &5895.92 &1.10   &0.76	&                      & Al {\sc ii}              	&3	&7042.06	&0.39	&0.64  	& \\
S {\sc ii}       &13   &5908.25	&0.05	&0.94  	&                      & Al {\sc ii}            	 &3 	&7056.60	&0.36	&0.67  	&\\
Si {\sc ii}    	 &8	   &5915.27	&0.10	&0.88  	&                      & Al {\sc ii}              	&3	&7063.64	&0.16	&0.81  	& \\
N {\sc ii}     	 &28   &5927.82	&0.04	&0.93 	&				       & He {\sc i}             	&10 	&7065.19	&0.01	&0.90  	&\\
N {\sc ii}       &28   &5931.79	&0.04	&0.93  	&                      & Fe {\sc ii}              	& 	&7067.44	&$-$0.08	&1.10	& dp   \\
N {\sc ii}       &28   &5940.25	&0.07	&0.92 	 &                     & $[$Fe {\sc ii}$]$          	&14 	&7155.14	&$-$0.58	&1.51	 & dp   \\
N {\sc ii}       &28   &5941.67	&0.10	&0.88 	 &                     & Ne {\sc i}	&6         	&7173.94	&0.16	&0.89  	& \\
Ne {\sc i}       &1	   &5944.83	&0.24	&0.73  	&                      & Fe {\sc ii}	&73        	&7222.39	&$-$0.22	&1.19	 & dp   \\
N {\sc ii}       &28   &5952.39	&0.04	&0.89	&                      & Fe {\sc ii}	&73        	&7224.51	&$-$0.62	&1.17	 & dp   \\
Si {\sc ii}      &4	   &5957.61	&0.18	&0.75	&                      & Ne {\sc i}	&3         	&7245.17	&0.27	&0.74  	&\\
Ne {\sc i}     	 &39   &5965.47	&0.14	&0.81	&                      & He {\sc i}	&45        	&7281.35	&0.18	&0.82  	& \\
Ne {\sc i}     	 &28   &5974.63	&0.18	&0.90	& +f                   & $[$Ca {\sc ii}$]$   	&1	   &7291.46	&$-$1.37	&2.19	 & dp   \\
Ne {\sc i}       &1	   &5975.53	&0.18	&0.82	& +p                   & $[$Fe {\sc ii}$]$          	&72 	&7301.57	&$-$0.19	&1.14	 & dp   \\
Si {\sc ii}      &4	   &5978.97	&0.21	&0.73	&                      & Fe {\sc ii}	&73        	&7307.97	&$-$0.67	&1.26	 & dp   \\
Fe {\sc ii}      &46   &5991.38	&$-$0.36&1.24	& dp                   & Fe {\sc ii}	&73        	&7310.24	&$-$0.11   	&1.20	 & dp   \\
S  {\sc ii}      &13   &5996.16	&0.08	&0.90	&                      & $[$Ca {\sc ii}$]$   	&1	   &7323.88	&$-$1.19	&2.02	 & dp   \\
Ne {\sc i}     	 &3	   &6030.00	&0.14	&0.77	&                      & Fe {\sc ii}	&          	&7376.46	&$-$0.31	&1.24	 & dp   \\
Ne {\sc i}       &3	   &6074.34	&0.28	&0.65	&                      & $[$Fe {\sc ii}$]$           	&14	&7388.16	&$-$0.26	&1.19	 & dp   \\
Fe {\sc ii}      &46   &6084.11	&$-$0.21&1.17	& dp                   & N  {\sc i}	&3              	&7423.63	&0.34   	&0.66 	&     \\
Fe {\sc ii}      &200  &6103.54	&$-$0.11&1.33	& dp                   & N  {\sc i}	&3         	&7442.28	&0.25	&0.71  	&     \\
Fe {\sc ii}      &46   &6113.33	&$-$0.16&1.08	& dp                   & Fe {\sc ii}	&73        	&7449.34	&$-$0.27	&1.23	 & dp   \\
Ne {\sc i}     	 &1	   &6143.06	&0.42	&0.56	&                      & $[$Fe {\sc ii}$]$           	&14	&7452.50	&$-$0.18	&1.21	 & dp   \\
Ne {\sc i}       &5	   &6163.59	&0.23	&0.56	&                      & Fe {\sc ii}	&73        	&7462.38	&$-$0.50	&1.27	 & dp   \\
DIB              &	   &6202.87	&0.29	&0.83	&                      & N  {\sc i}	&3        	 &7468.29	&0.38	&0.61  	&     \\
Ne {\sc i}       &1	   &6217.28	&0.21	&0.74	&                      & $[$Fe {\sc ii}$]$           	&72	&7479.70	&$-$0.13	&1.14	 & dp   \\
Al {\sc ii}      &10   &6226.18	&0.07	&0.90	&                      & Fe {\sc ii}	&73        	&7515.88	&$-$0.37	&1.25	 & dp   \\
Al {\sc ii}      &10   &6231.78	&0.15	&0.83	&                      & Fe {\sc ii}	&72        	&7533.42	&$-$0.19	&1.18	 & dp   \\
Al {\sc ii}      &10   &6243.36	&0.15	&0.83	&                      & Fe {\sc ii}	&73        	&7711.71	&$-$0.55	&1.27	 & dp   \\
Ne {\sc i}       &5	   &6266.50	&0.32	&0.64	&                      & O  {\sc i}	&1         	&7771.96	&0.31	&0.65  	&     \\
S {\sc ii}       &26   &6312.68	&0.16	&0.81	&                      & O  {\sc i}	&1         	&7774.18	&0.37	&0.67  	& +f    \\
Fe {\sc ii}      &	   &6317.38	&$-$0.18&1.11	& dp                   &  O  {\sc i}	&1         	&7775.40	&0.37	&0.74  	&  +p  \\  
Fe {\sc ii}      &	   &6331.97	&$-$0.09&1.09	&                      & He {\sc i}	&69        	&7816.16	&0.19	&0.82  	&     \\
Ne {\sc i}       &1	   &6334.43	&0.30	&0.64	&                      & $[$Fe {\sc ii}$]$          	&72 	&7841.40	&$-$0.12	&1.08	 & dp   \\
Si {\sc ii}      &2	   &6347.09	&0.42	&0.57	&                      & Fe {\sc ii}	&         	 &7866.55	&$-$0.12	&1.08	 & dp   \\
Fe {\sc ii}      &40   &6369.45	&$-$0.32&1.24	& dp                   & S {\sc ii}	&12        	&7967.43	&0.21	&0.80  	&     \\ 
Si {\sc ii}      &2	   &6371.36	&0.34	&0.59	&                      & Fe {\sc ii}	&          	&8048.28	&$-$0.18	&1.14	 & dp   \\
Ne {\sc i}       &3	   &6382.99	&0.28	&0.67	&                      & Ne {\sc i}	&23        	&8136.41	&0.12	&0.90  	&     \\
DIB              &	   &6397.39	&0.31	&0.77	&                      & N  {\sc i}	&2         	&8184.80	&0.26	&0.76  	&     \\
Ne {\sc i}     	 &1	   &6402.25	&0.53	&0.51	&                      & N  {\sc i}	&2         	&8187.95	&0.40	&0.66  	&     \\
Fe {\sc ii}      &74   &6407.30 &$-$0.30& 1.25  &                      & N  {\sc i}	&2         	&8210.64	&0.22	&0.78  	&     \\
DIB              &     &6413.90 &0.11 	&0.87	&                      & N  {\sc i}	&2         	&8216.28	&0.46	&0.63	  &     \\
Fe {\sc ii}      &74   &6416.91	&$-$0.38&1.22	& dp                   & N  {\sc i}	&2         	&8223.07	&0.34	&0.64  	&     \\
Fe {\sc ii}      &40   &6432.65	&$-$0.53&1.33	& dp                   & N  {\sc i}	&2         	&8242.34	&0.40	&0.60  	&     \\
Ne {\sc i}       &3	   &6506.53	&0.33	&0.66	&                      & Ne {\sc i}	&12        	&8300.33	&0.22	&0.83  	&     \\
Fe {\sc ii}      &40   &6516.05	&0.52	&1.38	& dp                   & S {\sc ii}	&12        	&8314.73	&0.28	&0.77  	&     \\
Ne {\sc i}     	 &	   &6532.88	&0.22	&0.77	&                      & He {\sc i}	&68        	&8361.77	&0.23	&0.85  	&     \\
H$\alpha$	     &1	   &6562.82	&$-$5.58&2.48	& dp                   & Ne {\sc i}	&12        	&8376.41	&0.35	&0.7  	& +f    \\
Ca {\sc i}       &1	   &6572.78 &$-$0.27&1.21	& dp                   & Ne {\sc i}	&12        	&8377.61	&0.35	&0.7  	& +p    \\
DIB              &     &6578.50	&0.23	&0.82	&                      & Ne {\sc i}	&18        	&8418.43	&0.21	&0.83  	&     \\
Ne {\sc i}       &6	   &6598.95	&0.23	&0.74	&                      & Fe {\sc ii}	&          	&8451.01	&$-$0.20	&1.12	 & dp    \\
N {\sc ii}     	 &31   &6610.58	&0.08	&0.91	&                      & Fe {\sc i}                	&	&8470.37	&$-$0.36	&1.20	 & dp    \\
N  {\sc i}       &20   &6653.46	&0.09	&0.90	&                      & Ca {\sc ii}	&2         	&8498.02	&$-$8.22	&3.91	 & dp    \\
\noalign{\smallskip}\hline\noalign{\smallskip}                         
\end{tabular}

\end{center}
\end{table*}

\setcounter{table}{3}

\begin{table*}
\begin{center}
\begin{center}
{\bf Table 4. continued: }{ List of lines identified in the spectrum of AS 386}
\end{center}

\begin{tabular}{lrcrrl|lrcrrl}
\hline\noalign{\smallskip}
Line         & Mult. &$\lambda_{\rm lab}$&EW & I/Ic & Comment &      Line         & Mult. &$\lambda_{\rm lab}$&EW & I/Ic & Comment \\  
\noalign{\smallskip}\hline\noalign{\smallskip}

Ca {\sc ii}	&2         	&8542.09	&$-$10.64	&3.97	 & dp    &                N  {\sc i}	&1         	&8747.35	&0.11	&0.85  	&     \\
N  {\sc i}	&8         	&8567.74	&0.10	&0.87  	&      & Ne {\sc i}	&27        	&8780.62	&0.24	&0.80  	&     \\
Ne {\sc i}	&30        	&8591.26	&0.13	&0.87  	&     & Ne {\sc i}	&38        	&8783.76	&0.16	&0.85  	&     \\
N  {\sc i}	&8         	&8594.01	&0.17	&0.83  	&     & Ne {\sc i}	&27        	&8853.87	&0.20	&0.85  	&     \\
$[$Fe {\sc ii}$]$          	&13 	&8616.96	&$-$0.25	&1.19	& dp    &Ne {\sc i}	&27       	 &8919.50	&0.04	&0.93  	&     \\
N  {\sc i}	&8         	&8629.24	&0.39	&0.92  	& +f    & Fe {\sc ii}	&          	&8926.66	&$-$0.40	&1.24	& dp    \\
Ne {\sc i}	&23        	&8634.65	&0.39	&0.84  	& +p    & N  {\sc i}	&15        	&9060.60	&0.77             	&0.74  	&  +f   \\
Ne {\sc i}	&33        	&8647.05	&0.60	&0.86  	&     & He {\sc i}	&77        	&9063.40	&0.77	&0.72  	& +p    \\
Ne {\sc i}	&33       	 &8654.38	&0.21	&0.80  	&     & He {\sc i}	&83        	&9210.28	&0.44	&0.78  	&     \\
Ca {\sc ii}	&2         	&8662.14	&$-$10.13	&4.03 	& dp    & Ne {\sc i}	&33        	&9300.85	&0.08	&0.83  	&     \\
N  {\sc i}	&1         	&8680.24	&0.42	&0.66  	&     & He {\sc i}	&67        	&9463.57	&0.25	&0.82 	 &     \\
Ne {\sc i}	&23        	&8681.92	&0.03	&0.95  	&     & He {\sc i}	&76        	&9516.51	&0.92	&0.56  	&     \\
N  {\sc i}	&1         	&8683.38	&0.23	&0.75  	&     & He {\sc i}	&75        	&9702.66	&0.32	&0.81  	&     \\
N  {\sc i}	&1        	 &8686.13	&0.29	&0.71  	&     & Fe {\sc ii}	&          	&9997.58	&$-$2.83	&1.71	 & dp    \\
N  {\sc i}	&1         	&8703.24	&0.42	&0.64 	 &     & He {\sc i}	&80        	&10072.10	&0.14	&0.88  	&     \\
N  {\sc i}	&1         	&8711.69	&0.28	&0.72  	&     & He {\sc i}	&89        	&10138.50	&0.67	&0.83  	&     \\
N  {\sc i}	&1         	&8718.82	&0.30	&0.70  	 &     & He {\sc i}	&74        	&10311.18	&0.94	&0.80  	&     \\
N  {\sc i}	&1         	&8728.88	&0.29	&0.79  	&     &&&&&&\\
\noalign{\smallskip}\hline\noalign{\smallskip}
\end{tabular}
\begin{tabular}{l}
Column information: \\ 
(1) -- element and ionization state, \\
(2) -- multiplet number (for the lines identified in Coluzzi 1993), \\
(3) --  line laboratory wavelength  in \AA, \\ 
(4) -- line EW in \AA\ (negative values are used for emission lines),  \\
(5) -- peak intensity of the line with respect to the local continuum (average of the two peaks for double-peaked lines), \\
 (6) -- comment on the line appearance in the spectrum or its ID. \\ 
 Abbreviations used: \\
 dp -- double-peaked (relative intensity of both peaks are shown); \\
+f -- blend with the following line; \\
 +p -- blend with the previous line  \\
 (for both +f and +p EW was measured for the entire blend).
\end{tabular}

\end{center}
\end{table*}

\end{document}